\documentclass[twoside]{article}
\usepackage{amssymb}
\usepackage{amsmath}

\usepackage[greek,english]{babel}
\languageattribute{greek}{polutoniko}
\usepackage{graphicx}

\newcommand{\be}{\begin{equation}}
\newcommand{\ee}{\end{equation}}
\newcommand{\beqar}{\begin{eqnarray}}
\newcommand{\eeqar}{\end{eqnarray}}
\newcommand{\Tr}{{\rm Tr}}
\newcommand{\half}{{\frac{1}{2}}}
\newcommand{\tr}{{\rm tr}}
\newcommand{\A}{{\rm A}}
\newcommand{\C}{{\rm C}}
\newcommand{\Ch}{{\hat {\rm C}}}
\newcommand{\F}{{\rm F}}
\newcommand{\Fh}{{\hat {\rm F}}}
\newcommand{\Sh}{{\hat S}}
\newcommand{\dd}{{\rm d}}
\newcommand{\D}{{\rm D}}
\newcommand{\thmn}{\theta^{\mu \nu}}
\newcommand{\thab}{\theta^{\alpha \beta}}
\newcommand{\tR}{{\hat R}}
\def \la {{\langle}}
\def \ra {{\rangle}}
\def \sp {_{\rm sp}}

\newcommand{\nc}{noncommutative\ }
\newcommand{\Nc}{Noncommutative\ }

\begin{document}

\title{Noncommutative Fluids}
\author{Alexios P. {\sc Polychronakos }\\ 
Physics Department, City College of New York\\
160 Convent Avenue, New York, NY 10031, USA}

\maketitle

\abstract{We review the connection between \nc gauge theory, matrix models and
fluid mechanical systems. The \nc Chern-Simons description of the quantum Hall
effect and bosonization of collective fermion states are used as specific examples.}

\vskip 0.2cm
\noindent
{\it To appear in the ``Bourbaphy'' S\'eminaire Poincar\'e X, Institut Henri Poincar\'e, Paris}

\tableofcontents

\section{Introduction}\label{sec1}

The idea that space may be a derived or emergent concept is a relatively old theme in
theoretical physics. In the context of quantum mechanics, observables are operators and
it is only their spectrum and mutual relations (commutators) that define their physical content.
Space, to the extent that it is observable, should be no different. The properties 
attributed to space from everyday experience -and postulated in newtonian mechanics and
special relativity- could be either exact or approximate, emerging in some particular
or partial classical limit. Other structures, extending or deforming the concepts of
classical geometry, and reducing to it under appropriate conditions, are conceivable.

This possibility has had an early emergence in speculations by Heisenberg himself.
It made reappearances in various guises and contexts \cite{Snyder}. One of the most strikingly
prescient of later developments in \nc gauge theory was the work of Eguchi and Kawai in
large-N single-plaquette lattice gauge theory \cite{EK}. It was, however, after the seminal and
celebrated work of Alain Connes that \nc geometry achieved the mathematical rigor and
conceptual richness that made it a major component of modern theoretical physics.
The concept made further inroads when it emerged as a property of spacetime solutions
derived from string theory \cite{CDS,SW} and, by now, it claims a huge body of research
literature.

One of the reasons that makes the idea of noncommutative spaces attractive is the
common language and connections that it provides between apparently disparate topics.
Indeed, as will be reviewed in this writeup, noncommutative physics unifies such
{\it a priori} different objects as gauge fields, membranes, fluids, matrix models and many-body
systems. (Some of the above connections can be established independently, but the full
continuum emerges only in the noncommutative setting.)

Unification of description usually brings unification of concepts. This raises the stakes
and elevates noncommutativity into a possibly fundamental property of nature. We could ask,
for instance, whether the eventual bringing together of gravity, quantum mechanics and
thermodynamics will arise out of some underlying fully noncommutative structure that shapes
into spacetime, quantum mechanics and statistical ensembles in some appropriate limit.
Whether this is indeed true is, of course, unclear and leaves room for wild speculation.

At this point, we should refrain from fantasizing any further and take a more pragmatic point
of view. The obvious question is: does noncommutativity buy us any advantage for physics as we
presently know it? It will
be the purpose of this expos\'e (as, I imagine, of the other talks in this session of the Poincar\'e
Institute) to demonstrate that this, indeed, is the case.

\section{Review of noncommutative spaces}

The concepts of noncommutative geometry will be covered by other speakers in this session
and there is probably little use in repeating them here. Moreover, there are many excellent and 
complete review articles, of which \cite{Harvey,DN,Szabo} are only a small sample.

Nevertheless, a brief summary will be presented here, for two main reasons.
Firstly, it will make this writeup essentially self-contained
and will minimize the need to refer to other sources for a coherent reading; and secondly,
the level and tone of the presentation will be adapted to our needs, and hopefully will serve
as a low-key alternative to more rigorous and complete treatments.

\subsection{The operator formulation}

The simplest starting point for the definition of noncommutative spaces is through the
definition of noncommutative coordinates. This is the approach that is most closely
related to physics, making the allusions to quantum mechanics most explicit, and is
therefore also the most common one in physics texts. In this, the noncommutative spaces
are defined in terms of their coordinates $x^\mu$, which are abstracted into (linear) operators.
Such coordinates can be added and multiplied (associatively), forming a full operator algebra,
but are not (necessarily) commutative. Instead, they obey the commutation relations
\be
[ x^\mu , x^\nu ] = i \theta^{\mu \nu} ~,~~~\mu,\nu=1, \dots d
\label{xx}
\ee
The antisymmetric two-tensor $\theta^{\mu \nu}$ could be itself an operator, but is usually
taken to commute with all $x^\mu$ (for `flat' noncommutative spaces) and is, thus, a set of
ordinary, constant c-numbers. Its inverse, when it exists,
\be
\omega_{\mu \nu} = (\theta^{-1} )_{\mu \nu}
\ee
defines a constant two-form $\omega$ characterizing the 
noncommutativity of the space.

Clearly the form of $\theta$ can be changed by redefining the coordinates of the space.
Linear redefinitions of the $x^\mu$, in particular, would leave $\thmn$ a
c-number (nonlinear redefinitions will be examined later).
We can take advantage of this to give a simple form to $\thmn$. Specifically, by
an orthogonal transformation of the $x^\mu$ we can bring $\thmn$ to a Darboux
form consisting of two-dimensional blocks proportional to $i\sigma_2$ plus a set of
zero eigenvalues. This would decompose the space into a direct sum of mutually commuting
two-dimensional noncommutative subspaces, plus possibly a number of commuting coordinates
(odd-dimensional  spaces necessarily have at least one commuting coordinate).
In general, there will be $2n$ properly noncommuting coordinates $x^\alpha$ ($\alpha
= 1, \dots 2n$) and $q=d-2n$ commuting ones $Y^i$ ($i=1,\dots q$). 
In that case $\omega$ will be defined 
as the inverse of the projection $\bar \theta$ of $\theta$ on the fully
noncommuting subspace:
\be
\omega_{\alpha \beta} = ({\bar \theta}^{-1} )_{\alpha \beta} ~,~~~
\omega_{ij} = 0
\ee

The actual \nc space can be though of as a representation of the above operator algebra (\ref{xx}),
acting on a set of states. For real spaces the operators $x^\mu$ will be considered
hermitian, their eigenvalues corresponding to possible values of the
corresponding coordinate. Not all coordinates can be diagonalized simultaneously, so
the notion of `points' (sets of values for all coordinates $x^\mu$) is absent. The
analogy with quantum mechanical coordinate and momentum is clear, with each `Darboux'
pair of noncommutative coordinates being the analog of a canonical quantum pair. Nevertheless,
a full set of geometric notions survives, in particular relating to fields on the
space, as will become clear.

The representation of $x^\mu$ can be reducible or irreducible. For the commuting components
$Y^i$ any useful representation must necessarily be reducible, else the corresponding
directions would effectively be absent (consisting of a single point). 
States are labeled by the values of these coordinates $y^i$, taken to be 
continuous. The rest of the space, consisting of canonical Heisenberg pairs,
admits the tensor product of Heisenberg-Fock Hilbert spaces 
(one for each two-dimensional noncommuting subspace $k=1, \dots n$) as its
unique irreducible representation. In general, we can have a reducible representation
consisting of the direct sum of $N$ such irreducible components
for each set of values $y^i$, labeled by an extra index $a=1,\dots N$
(we shall take $N$ not to depend on $y^i$). A complete basis for the states,
then, can be
\be
|n_1 , \dots n_n ; y^1 , \dots y^q ; a \ra
\ee
where $n_k$ is the Fock (oscillator) excitation number of the $k$-th 
two-dimensional subspace.

Due to the reducibility of the above representation, the operators
$x^\mu$ do not constitute a complete set. To make the set complete,
additional operators need be introduced. To deal with the reducibility
due to the values $y^i$, we consider translation (derivative)
operators $\partial_\mu$. These are defined through their
action on $x^\mu$, generating constant shifts:
\be
[\partial_\mu , x^\nu ] = \delta_\mu^\nu
\ee
On the fully noncommutative subspace these are inner automorphisms
generated by
\be
\partial_\alpha = -i \omega_{\alpha \beta} x^\beta
\ee
For the commutative coordinates, however, extra operators have to be
appended, shifting the Casimirs $Y^i$ and thus acting on the 
coordinates $y^i$ as usual derivatives.

To deal with the reducibility
due to the components $a = 1, \dots N$, we need to introduce yet another set
of operators in the full representation space mixing the above $N$ components.
Such a set are the hermitian $U(N)$ operators $G^r$, $r=1, \dots N^2$ that
commute with the $x^\mu$, $\partial_\mu$ and mix the components $a$.
(We could, of course, choose these operators to be the $SU(N)$ subset, eliminating
the trivial identity operator.)
The set of operators $x^\alpha , \partial_i , G^r$ is now complete.

Within the above setting, we can define field theories on a noncommutative
space. Fields are the analogs of functions of coordinates $x^\mu$; that is,
arbitrary operators in the universal enveloping algebra of the $x^\mu$. 
In general, the above fields
are {\it not} arbitrary operators on the full representation space, since they
commute with $\partial_i$ and $G^r$. In particular, they act `pointwise' on the
commutative coordinates $Y^i$ are are, therefore, ordinary functions of the $y^i$.

We can, of course, define fields depending also on the remaining operators.
Fields involving operators $G^r$ are useful, as they act as 
$N \times N$ matrices on components $a$.
They are the analogs of matrix-valued fields and will be useful in constructing
gauge theories. We could further define operators
that depend on the commutative derivatives $\partial_i$. These have no
commutative analog, and will not be considered here. Notice, however, that on
fully noncommutative spaces (even-dimensional spaces without commutative
components), the matrix-valued fields $f^{ab} ( x^\mu )$ constitute the {\it full}
set of operators acting on the representation space.

The fundamental notions completing the discussion of noncommutative field theory are
the definitions of derivatives and space integral. Derivatives of a function $f$ are
defined as commutators with the corresponding operator:
\be
\partial_\mu \cdot f = [\partial_\mu , f]
\ee
That is, through the adjoin action of the operator $\partial_\mu$ on fields (we use the
dot to denote this action). For the commutative derivatives $\partial_i$ this is the
ordinary partial derivative $\partial / \partial_{y^i}$. For the noncommutative coordinates,
however, such action is generated by the $x^\alpha$ themselves, as $\partial_\alpha = -i 
\omega_{\alpha \beta} x^\beta$. So the notion of coordinates and derivatives on purely
noncommutative spaces fuses, the distinction made only upon specifying the action of
the operators $x^\alpha$ on fields (left- or right- multiplication, or adjoin action).

The integral over space is defined as the trace in the representation space,
normalized as:
\be
\int d^d x = \int d^q y  ~ \tr'  ~\sqrt{\det (2\pi\theta )} \, \tr \equiv \Tr  
\ee
where $\tr$ is the trace over the Fock spaces and $\tr'$ is the trace
over the degeneracy index $a=1,\dots N$. This corresponds to a space integral
and a trace over the matrix indices $a$. The extra determinant factor ensures the
recovery of the proper commutative limit (think of semiclassical quantization, or
the transition from quantum to classical statistical mechanical partition functions.)

All manipulations within ordinary field theory can be transposed here, with a
noncommutative twist. For instance, the fact that the integral of a total derivative
vanishes (under proper boundary conditions), translates to the statement
that the trace of a commutator vanishes, and its violation by fields with
nontrivial behavior at infinity is mirrored in the nonvanishing trace of the
commutator of unbounded, non-trace class operators, such as the noncommutative
coordinates themselves. Finite-dimensional truncations of the above coordinate-derivative
operators can be used for numerical simulations of \nc field theories on the basis
of the above formulae \cite{numerics}.

\subsection{Weyl maps, Wigner functions and $*$-products}

The product of noncommutative fields is simply the product
of the corresponding operators, which is clearly associative but not commutative.
It is also not `pointwise', as the
notion of points does not even exist. Nevertheless, in the limit
$\thmn \to 0$ we recover the usual (commutative) geometry and algebra of functions.
Points are recovered as any set of states whose spread $\Delta x^\mu$ in each coordinate
$x^\mu$ goes to zero in the commutative limit. Such a useful set is, e.g., the set
of coherent states in each noncommutative (Darboux) pair of coordinates with
average values $x^\mu$.

Observations like that can form the basis of a complete
mapping between noncommutative fields and commutative functions $f(x)$,
leading to the notion of the `symbol' of $f(x)$ and the star-product.
Specifically, by expressing fields as functions of the fundamental operators $x^\mu$ and
ordering the various $x^\mu$ in the expressions for 
the fields in a prescribed way, using their known commutators, establishes a one-to-one
correspondence between functions of operators and ordinary functions. This is reminiscent of,
and in fact equivalent to, the Wigner function mapping of a quantum mechanical operator
onto the classical phase space (see \cite{Zachos} for a simple review).

The ordering that is most usually adopted is the fully symmetric Weyl
ordering, in which monomials in the $x^\mu$ are fully symmetrized. It is simplest to
work with the Fourier transforms of functions, since exponentials of linear combinations
of $x^\mu$ are automatically Weyl ordered. So a classical function $f(x)$, with Fourier
transform ${\tilde f}(k)$, is mapped to the operator (\nc field) $f$ as:
\be
f = \int dk \, e^{i k_\mu x^\mu} {\tilde f}(k)
\ee
(the integral over $k$ is of the appropriate dimensionality). Conversely, the `symbol'
(commutative function) corresponding to an operator $f$ can be expressed as:
\be
{\tilde f}(k) = \sqrt{\det (\theta/2\pi) } \, \tr \, f e^{-i k_\mu x^\mu}
\ee
where the above trace is taken over an irreducible representation of the \nc coordinates.
This reproduces scalar functions. For matrix-valued \nc fields $f$, acting nontrivially
on a direct sum of $N$ copies of the irreducible representation, the above expression
generalizes to
\be
{\tilde f}^{ab}(k) = \sqrt{\det (\theta/2\pi) } \sum_n 
\la n,a | f e^{-i k_\mu x^\mu} |n,b \ra
\ee
where $|n,a \ra$ are a complete set of states for the $a$-th copy of the irreducible
representation, reproducing a matrix function of commutative variables. Hermitian operators
$f$ map to hermitian matrix functions $f^{ab} (x)$ or, in the case $N=1$, real functions.

On can show that, under the above mapping, derivatives and integrals of \nc fields
map to the standard commutative ones for their symbol.
The product of operators, however, maps to a new function, called the star-product of 
the corresponding functions \cite{star}:
\be
f \leftrightarrow f(x) ~,~~g \leftrightarrow g(x) ~~~\implies ~~~f g \leftrightarrow (f * g)(x)
\ee
The star product can be written explicitly in terms of the Fourier transforms of functions as
\be
(f*g)(k) = \int dk \, {\tilde f} (q) \, {\tilde g} (k-q) \, 
e^{\frac{i}{2} \theta^{\mu \nu} k_\mu k_\nu}
\ee
This is the standard convolution of Fourier transforms, but with an extra phase factor.
The resulting $*$-product is associative but \nc and also nonlocal in the coordinates $x^\mu$.
The commutator of two \nc fields maps to the so-called star, or Moyal, brackets of their symbols.

The above mapping has the
advantage that it circumnavigates the conceptual problems of noncommutative
geometry by working with familiar objects such as ordinary functions and their
integral and derivatives, trading the effects of noncommutativity for a nonlocal,
noncommutative function product. It can, however, obscure the beauty and conceptual
unification that arises from noncommutativity and make some issues or calculations
unwieldy. In what follows, we shall stick with the operator formulation as exposed
above. Translation into the $*$-product language can always be done at any desired stage.

\section{Noncommutative gauge theory}

Gauge theory on noncommutative spaces becomes particularly attractive \cite{MSSW,Klimcik,KaWa}.
Gauge fields $A_\mu$ are hermitian operators acting on the representation space.
Since they do not depend on $\partial_i$
they cannot shift the values of $y^i$, while they act nontrivially on the
fully noncommuting subspace. They have effectively become big matrices
acting on the full Fock space with elements depending on the commuting
coordinates. Derivatives of these fields are defined through the
adjoin action of $\partial_\mu$
\be
\partial_\mu \cdot A_\nu = [ \partial_\mu , A_\nu ]
\ee

Using the above formalism, gauge field theory can be built in a way
analogous to the commuting case. Gauge transformations are unitary
transformations in the full representation space. Restricting $A_\mu$ to 
depend on the coordinates only, as above, produces the so-called $U(1)$
gauge theory. $U(N)$ gauge theory can be obtained by relaxing this
restriction and allowing $A_\mu$ to also be a function of the $G^r$ and
thus act on the index $a$.

\subsection{Background-independent formulation}

The basic moral of the previous section is that noncommutative gauge theory
can be written in a universal way \cite{APA,GN,GHI}. 
In the operator formulation no special distinction 
needs be done between $U(1)$ and $U(N)$ theories, nor need gauge and spacetime
degrees of freedom be treated distinctly. The fundamental operators of the theory are
\be
D_\mu = -i\partial_\mu + A_\mu
\ee
corresponding to covariant derivatives. Gauge transformations are simply unitary
conjugations of the covariant derivative operators by a unitary field $U$. That is,
the $D_\mu$ transform covariantly: 
\be
D_\mu \to U^{-1} D_\mu U
\ee
This reproduces the (noncommutative version of the) standard gauge transformation of $A_\mu$:
\be
A_\mu \to -i U^{-1} \partial_\mu \cdot U + U^{-1} A_\mu U
\ee
For the fully noncommutative components, covariant derivative operators assume the form
\be
D_\alpha = \omega_{\alpha \beta} x^\beta + A_\alpha = \omega_{\alpha \beta} ( x^\beta + 
\theta^{\beta \gamma} A_\gamma ) = \omega_{\alpha \beta} X^\beta
\ee

The above rewriting is important in various ways. It stresses the fact that, on fully
noncommutative spaces, the separation of $D_\alpha$ into $x^\alpha$ (coordinate) and $A_\alpha$
(gauge) is largely arbitrary and artificial: both are operators acting on the Hilbert space
on an equal footing, the distinction between `derivative' and `coordinate' having been eliminated.
This separation is also gauge dependent, since a unitary transformation will mix the two parts.
In effect, gauge transformations mix spatial and gauge degrees of freedom! Further, it is not
consistent any more to consider strictly $SU(N)$ gauge fields. Even if $A_\mu$ is originally
traceless in the $N$-dimensional index $a$, gauge transformations $U$ cannot meaningfully be
restricted to $SU(N)$: the notion of partial trace of an operator with respect to one component of
a direct product space makes sense, but the notion of partial determinant does not. A gauge
transformation will always generate a $U(1)$ part for $A_\mu$, making $U(N)$ gauge theory the only
theory that arises naturally.

The above rewriting also introduces the `covariant coordinate' field $X^\alpha$ that combines the
ordinary coordinate and gauge fields in a covariant way and is dual to the covariant derivative.
\Nc gauge theory can be constructed entirely in terms of the $X^\alpha$. These, in turn, can be thought
of as `deformed' coordinates, the deformation being generated by (the dual of) gauge fields, which
alludes to stretching membranes and fluids. All this is relevant in the upcoming story.

Any lagrangian built entirely out of $D_\mu$ will lead to a gauge invariant action, 
since the trace will remain invariant under any unitary transformation. The standard
Maxwell-Yang-Mills action is built by defining the field strength
\be
F_{\mu \nu} = \partial_\mu \cdot A_\nu - \partial_\nu \cdot A_\mu + i
[ A_\mu , A_\nu ] = i [ D_\mu , D_\nu ] - \omega_{\mu \nu}
\ee
and writing the standard action
\be
S_{LYM} = \frac{1}{4g^2} \Tr F_{\mu \nu} F^{\mu \nu} = - \frac{1}{4g^2} \Tr
 ([ D_\mu , D_\nu ] +i \omega_{\mu \nu} )^2
\ee
where $\Tr$ also includes integration over commutative components $y^i$.
In the above we used some c-number metric tensor $g^{\mu \nu}$ to raise the indices of $F$. 
Note that
the operators $\partial_\alpha \cdot$, understood to act in the adjoin on fields, commute,
while the operators $\partial_\alpha = -i \omega_{\alpha \beta} X^\beta$ have a nonzero
commutator equal to 
\be
[\partial_\alpha , \partial_\beta ] = i \omega_{\alpha \beta}
\ee
This explains the extra $\omega$-term appearing
in the definition of $F$ in terms of covariant derivative commutators.

One can, however, just as well work with the action
\be
{\hat S}_{LYM} = \frac{1}{4g^2} \Tr {\hat F}_{\mu \nu} {\hat F}^{\mu \nu} = 
- \frac{1}{4g^2} \Tr [ D_\mu , D_\nu ] [ D^\mu , D^\nu ]
\label{DDD}
\ee
Indeed, $\hat S$ differs from $S$ by a term proportional to $\Tr \omega^2$, which is an
irrelevant (infinite) constant, as well as a term proportional to $\omega^{\mu \nu}
\Tr [ D_\mu , D_\nu ]$, which, being the trace of a commutator (a `total derivative'),
does not contribute to the equations of motion. The two actions lead to the same
classical theory. Note that $\thmn$ or $\omega_{\mu \nu}$ do not appear in the action.
These quantities arise only in the commutator of \nc coordinates. Since the $x^\mu$ do not
explicitly appear in the action either (being just a gauge-dependent part of $D_\mu$),
all reference to the specific \nc space has been eliminated! This is the `background
independent' formulation of \nc gauge theory that stresses its universality.

\subsection{Superselection of the \nc vacuum}

How does, then, a particular \nc space arise in this theory?
The equations of motion for the operators $D_\mu$ are
\be
[D^\mu , [ D_\mu , D_\nu ]] = 0
\ee
The general operator solution of this equation is not fully known.
Apart from the trivial solution $D_\mu = 0$, it admits as solution all 
operators with c-number commutators, satisfying
\be
[ D_\mu , D_\nu ] = -i\omega_{\mu \nu}
\ee
for some $\omega$. This is the classical `noncommutative vacuum', where 
$D_\mu = -i\partial_\mu$, and expanding $D_\mu$ around this vacuum leads 
to a specific noncommutative gauge theory.

Quantum mechanically, $\omega_{\mu \nu}$ are 
superselection parameters and the above vacuum is stable.  To see this, 
assume that the time direction is commutative and consider the collective mode
\be
D_\alpha = -i\lambda_{\alpha \beta} \partial_\beta
\ee
with $\lambda_{\alpha \beta}$ parameters depending only on time.
This mode would change the noncommutative vacuum while leaving the gauge
field part of $D_\alpha$ unexcited. $\omega$ gets modified into
\be
\omega_{\mu \nu}' = \lambda_{\mu \alpha} \omega_{\alpha \beta} 
\lambda_{\beta \nu}
\ee
The action implies a quartic potential for this mode, with a strength proportional
to $\Tr 1$, and a kinetic term proportional to $\Tr \partial_\alpha \partial_\beta$.
(There is also a gauge constraint which does not alter the qualitative dynamical
behavior of $\lambda$.) Both potential and kinetic terms are infinite, 
and to regularize them we should truncate each Fock space
trace up to some highest state $\Lambda$, corresponding to a finite volume 
regularization (the area of each noncommutative two-dimensional subspace has 
effectively become $\Lambda$). One can check that the potential term would 
grow as $\Lambda^n$ while the kinetic term
would grow as $\Lambda^{n+1}$. Thus the kinetic term dominates; the above 
collective degrees of freedom acquire
an infinite mass and will remain ``frozen'' to whatever initial value
they are placed, in spite of the nontrivial potential. (This is analogous to
the $\theta$-angle of the vacuum of four-dimensional nonabelian gauge theories:
the vacuum energy depends on $\theta$ which is still superselected.)
Quantum mechanically there is no interference between different
values of $\lambda$ and we can fix them to some c-number value, thus fixing
the noncommutativity of space \cite{APB}. This phenomenon is similar to symmetry breaking,
but with the important difference that the potential is not flat along changes of the
``broken'' vacuum, and consequently there are no Goldstone bosons.

In conclusion, we can start with the action (\ref{DDD}) as the definition of our
theory, where $D_\mu$ are arbitrary operators (matrices) in some space. Gauge
theory is then defined as a perturbation around a (stable) classical vacuum.
Particular choices of this vacuum will lead to standard noncommutative gauge
theory, with $\theta^{\mu \nu}$ and $N$ appearing as vacuum parameters.
Living in any specific space and gauge group amounts to landscaping!

\subsection{Noncommutative Chern-Simons action}

A particularly useful and important type of action in gauge theory is the Chern-Simons
term \cite{DJT}. This is a topological action, best written in terms of differential forms.
In the commutative case, we define the one- and two-forms
\be
\A = i A_\mu dx^\mu ~,~~~\F = \dd \A + \A^2 = \frac{i}{2} \Bigl( \partial_\mu 
A_\nu - \partial_\nu A_\mu + i [A_\mu , A_\nu ] \Bigr) dx^\mu dx^\nu 
\ee
The Chern-Simons action $S_{2n+1}$ is the integral of the $2n+1$-form 
$\C_{2n+1}$ satisfying
\be
\dd \C_{2n+1} = \tr \F^{n+1}
\label{dCF}
\ee
By virtue of (\ref{dCF}) and the gauge invariance of $\tr \F^n$
it follows that $S_{2n+1}$ is gauge invariant up to total derivatives,
since, if $\delta$ stands for an infinitesimal gauge transformation,
\be
\dd \delta \C_{2n+1} = \delta \dd \C_{2n+1} = \delta \tr \F^n = 0 ~,
~~{\rm so}~~~ \delta C_{2n+1} = \dd \Omega_{2n}
\ee
The integrated action is therefore invariant under infinitesimal
gauge transformations. Large gauge transformations may lead to 
an additive change in the action and they usually imply a quantization
of its coefficient \cite{DJT,Alvarez}.
As a result, the equations of motion derived from
this action are gauge covariant and read
\be
\frac{\delta S_{2n+1}}{\delta \A} = \frac{\delta}{\delta \A} 
\int \C_{2n+1} = (n+1) \F^n 
\label{defC}
\ee
The above can be considered as the defining relation for $\C_{2n+1}$.

We can define corresponding \nc Chern-Simons actions \cite{CF}-\cite{BLP}. 
To this end, we shall adopt the differential form language \cite{APB} and define the
usual basis of one-forms $dx^\mu$ as a set of formal anticommuting parameters
with the property 
\be
dx^\mu dx^\nu = - dx^\nu dx^\mu ~,~~~
dx^{\mu_1} \cdots dx^{\mu_d} = \epsilon^{\mu_1 \dots \mu_d}
\ee
Topological actions do not involve the metric tensor and can be written as 
integrals of $d$-forms. The only dynamical objects available in \nc gauge theory 
are $D_\mu$ and thus the only form that we can write is
\be
\D = i dx^\mu D_\mu = \dd + \A
\ee
where we defined the exterior derivative and gauge field one-forms
\be
\dd = dx^\mu \partial_\mu ~,~~~ \A = i dx^\mu A_\mu 
\ee
(note that both $\D$ and $\A$ as defined above are antihermitian). 
The action of the exterior derivative $\dd$ on an operator $p$-form $H$,
$\dd \cdot {\rm H}$, 
yields the $p+1$-form $dx^\mu [\partial_\mu , {\rm H}]$ and is given by
\be
\dd \cdot {\rm H} = \dd {\rm H} - (-)^p {\rm H} \dd
\ee
In particular, on the gauge field one-form $\A$ it acts as 
\be
\dd \cdot \A = \dd \A + \A \dd
\ee
Correspondingly, the covariant exterior derivative of $\rm H$ is
\be
\D \cdot {\rm H} = \D {\rm H} - (-)^p {\rm H} \D
\ee
As a result of the noncommutativity of the operators $\partial_\mu$,
the exterior derivative operator is not nilpotent but rather satisfies
\be
\dd^2 = \omega ~,~~~~ \omega = \frac{i}{2} dx^\mu dx^\nu \omega_{\mu \nu}
\ee
We stress, however, that $\dd \cdot$ is still nilpotent since $\omega$
commutes with all operator forms:
\be
\dd \cdot \dd \cdot {\rm H} = [\dd , [\dd , {\rm H}]_\mp ]_\pm =
\pm [\omega , {\rm H} ] = 0
\ee
The two-form $\Fh = \frac{i}{2}dx^\mu dx^\nu {\hat F}_{\mu \nu}$ is simply
\be
\Fh = \D^2 = \frac{1}{2} \D \cdot \D = 
\omega + \dd \A + \A \dd + \A^2 = \omega + \F
\ee
where $\F = \frac{i}{2} dx^\mu dx^\nu F_{\mu \nu}$ is the conventionally
defined field strength two-form.

The most general $d$-form that we can write involves arbitrary combinations of
$\D$ and $\omega$. If, however, we adopt the view that $\omega$ should arise as
a superselection (vacuum) parameter and not as a term in the action, the unique
form that we can write is $\D^d$ and the unique action
\be
\Sh_d = \frac{d+1}{2d} \Tr \, \D^d = \Tr \, \C_d
\label{SCS}
\ee
This is the Chern-Simons action. The coefficient was chosen to conform with
the commutative definition, as will be discussed shortly. In even dimensions 
$\Sh_d$ reduces to 
the trace of a commutator $\Tr [\D , \D^{d-1} ]$, a total derivative that 
does not affect the equations of motion and corresponds to a topological term.
In odd dimensions it becomes a nontrivial action.

$\Sh_d$ is by construction gauge invariant. To see that it also 
satisfies the defining property of a Chern-Simons form (\ref{defC}) 
is almost immediate: $\delta / \delta \A = \delta / \delta \D$ and thus,
for $d=2n+1$:
\be
\frac{\delta}{\delta \A} \Tr \, \D^{2n+1} = (2n+1) \, \D^{2n} = (2n+1) \, \Fh^n
\ee
So, with the chosen normalization in (\ref{SCS}) we have the defining condition
(\ref{defC}) with $\Fh$ in the place of $\F$. What is less obvious is that $\Sh_D$
can be written entirely in terms of  $\F$ and $\A$ and that, for commutative spaces,
it reduces to the standard Chern-Simons action. To achieve that, one must expand
$\C_D$ in terms of $\dd$ and $\A$, make use of the cyclicity of trace and the condition
$\dd^2 = \omega$ and reduce the expressions into ones containing 
$\dd \A + \A \dd$ 
rather than isolated $\dd$ s.  The condition 
\be
\Tr \omega^n \dd = 0 
\ee
which is a result of the fact that $\partial_\mu$ is off-diagonal for both 
commuting and noncommuting dimensions, can also be used to get rid of overall 
constants. This
is a rather involved procedure for which we have no algorithmic approach.
(Specific cases will be worked out later.) Note,
further, that the use of the cyclicity of trace implies that we dismiss total derivative
terms (traces of commutators). Such terms do not affect the equations of motion.
For $d=1$ the result is simply
\be
\Sh_1 = \Tr \A
\ee
which is the `abelian' one-dimensional Chern-Simons term. 
For $d=3$ we obtain
\be
\Sh_3 = \Tr (\A \F - \frac{1}{3} \A^3 ) + 2 \Tr (\omega \A )
\ee
where we used the fact that $\Tr [\A (\dd \A + \A \dd )] = 2\Tr (\A^2 \dd)$.  The first
term is the noncommutative version of the standard three-dimensional 
Chern-Simons term, while the second is a lower-dimensional Chern-Simons 
term involving explicitly $\omega$.

We can get the general expression for $\Sh_d$ by referring to the defining
relation. This reads
\be
\frac{\delta}{\delta \A} \Sh_{2n+1} = (n+1) \Fh^n = (n+1) (\F +\omega)^n
= (n+1) \sum_{k=0}^n {n \choose k} \omega^{n-k} \F^k 
\ee
and by expressing $\F^k$ as the $\A$-derivative of the standard Chern-Simons
action $S_{2k+1}$ we get
\be
\frac{\delta}{\delta \A}  \left\{ \Sh_{2n+1}  - \sum_{k=0}^n 
{n+1 \choose k+1}  \omega^{n-k}  S_{2k+1} \right\} = 0
\ee
So the expression in brackets must be a constant, easily seen to be zero by
setting $\A=0$. We therefore have
\be
\Sh_{2n+1}  = \sum_{k=0}^n {n+1 \choose k+1}  \Tr \omega^{n-k}  \C_{2k+1} 
\label{SSh}
\ee
We observe that we get the $2n+1$-dimensional Chern-Simons action plus
all lower-dimensional actions with tensors $\omega$ inserted to complete the
dimensions. Each term is separately gauge invariant and we could have chosen
to omit them, or include them with different coefficient. It is the specific combination
above, however, that has the property that it can be reformulated in a way that does
not involve $\omega$ explicitly. The standard Chern-Simons action can also
be written in terms of $\D$ alone by inverting (\ref{SSh}):
\be
S_{2n+1} = (n+1) \Tr \int_0^1 \D (t^2 \D^2 -\omega)^n dt =  \Tr \sum_{k=0}^n 
{n+1 \choose k+1} \frac{k+1}{2k+1}  (-\omega)^{n-k} \D^{2k+1}
\label{ShS}
\ee
For example, the simplest nontrivial \nc action in 2+1 dimensions reads
\be
S_3 = \Tr \left( \frac{2}{3} \D^3 - 2 \omega \D \right)
\ee
The above can be written more explicitly in terms of the two spatial covariant
derivatives $D_{1,2}$, which are operators acting on the \nc space, and the
temporal covariant derivative $D_0 = dt (\partial_t + i A_0 )$, which contains
a proper derivative operator in the commutative direction $x^0 = t$ and a \nc
gauge field $A_0$:
\be
S_3 = \int dt\, 2\pi \theta\, \Tr \left\{ \epsilon^{ij} ({\dot D}_i
+i [A_0 , D_i ]) D_j + \frac{2}{\theta} A_0 \right\}
\label{CSDD}
\ee
Note that the overall coefficient of the last, linear term is independent of $\theta$.

We also point out a peculiar property of the Chern-Simons form $\Ch_{2n+1}$.
Its covariant derivative yields $\Fh^{n+1}$:
\be
\D \cdot \Ch_{2n+1} = \D \Ch_{2n+1} + \Ch_{2n+1} \D = \frac{2n+2}{2n+1}
\, \Fh^{n+1}
\label{DCF}
\ee
A similar relation holds between $\C_d$ (understood as the form appearing inside the
trace in the right hand side of (\ref{ShS})) and $\F$. Clearly the standard Chern-Simons 
form does not share this property.  Our $C_d$
differs from the standard one by commutators that cannot all be written as ordinary
derivatives (such as, {\it e.g.}, $[\dd , \dd \A]$). These unconventional terms turn
$C_d$ into a covariant quantity that satisfies (\ref{DCF}).

\subsection{Level quantization for the \nc Chern-Simons action}

We conclude our consideration of the \nc Chern-Simons action by considering the 
quantization requirements for its coefficient \cite{NPq,BLP}.

In the commutative case, a quantization condition for the
coefficient of nonabelian Chern-Simons actions (`level quantization')
is required for global gauge invariance.
This has its roots in the topology of the group of gauge transformations in the given
manifold. E.g., for the 3-dimensional term, the fact that $\pi_3 [SU(N)] = Z$ for any
$N>1$ implies the existence of topologically nontrivial gauge transformations and
corresponding level quantization.

For the \nc actions we have not studied the topology of the gauge group. This would appear
to be a hard question for a `fuzzy' \nc space, but in fact is is well-defined and easy
to answer: gauge transformations are simply unitary transformations on the full representation
space on which $X^\mu$ or $D_\mu$ act. This space is infinite dimensional, so we are dealing
with (some version of) $U(\infty)$. Two observations, however, elucidate the answer.
First, for odd-dimensional \nc spaces there is always one (and in general only one) commutative
dimension $t$, conventionally called time and compactified to a circle;
and second, if we require gauge transformations to
act trivially at infinity, we are essentially restricting the corresponding unitary operators
to have finite support on the representation space and be bounded. So the relevant gauge
transformations are essentially $U(N)$ matrices of the form $U (t)$, where $N$ is the
`support' of $U$, that is, the dimension
of the subspace of the Hilbert space on which $U$ acts nontrivially. The relevant topology
is $S^1 \to U(N)$ and is nontrivial due to the $U(1)$ factor in $U(N)$:
\be
\pi_1 [U(N)] = \pi_1 [U(1)] = Z
\ee
This is true for {\it any} \nc gauge theory, abelian or nonabelian. A `winding number one'
transformation would be a matrix of the form
\be
U (t) = e^{i \frac{2\pi}{N} t} {\tilde U} (t) ~,~~~ t \in [0,1]
\ee
with ${\tilde U}$ an $SU(N)$ matrix satisfying ${\tilde U}(0)=1$ and ${\tilde U}(1)
= {\rm exp}(-i \frac{2\pi}{N})$, a $Z_N$ matrix. This satisfies $U(0) = U(1) = 1$ but cannot be
smoothly deformed to $U(t)=1$.

What is the change, if any, of the \nc Chern-Simons action under the above transformation?
We may look at the explicit form (\ref{CSDD}) of $S_3$ to decide it. The first, cubic term
is completely gauge invariant. Indeed, under a gauge transformation the quantity inside the
trace and integral transforms covariantly
\be
\epsilon^{ij} ({\dot D}_i +i [A_0 , D_i ]) D_j \to
U(t)^{-1} \left[ \epsilon^{ij} ({\dot D}_i +i [A_0 , D_i ]) D_j \right] U(t)
\ee
and upon tracing it remains invariant. The term $A_0$, however, transforms as
\be
A_0 \to U(t)^{-1} A_0 U(t) -i U(t)^{-1} {\dot U}(t)
\ee
The last term gives a nontrivial contribution to the action equal to
\be
\Delta S_3 = -i 4\pi \, \int_0^1 dt\, \tr U(t)^{-1} {\dot U}(t)
\ee
The $SU(N)$ part $\tilde U$ of $U(t)$ does not contribute in the above, since
${\tilde U}^{-1} {\dot {\tilde U}}$ is traceless. The $U(1)$ factor, however, 
contributes a part equal to
\be
\Delta S_3 = -i 4\pi \, \int_0^1 dt\, i \frac{2\pi}{N} \tr 1 = 8 \pi^2
\ee
The coefficient of the action $\lambda$ should be such that the overall change of the
action be 
quantum mechanically invisible, that is, a multiple of $2\pi$. We get
\be
\lambda \, 8 \pi^2 = 2\pi n ~~~{\rm or}~~~ \lambda = \frac{n}{4\pi}
\ee
with $n$ an integer.

The above quantization condition is independent of $\theta$ and conforms with the level
quantization of the commutative nonabelian Chern-Simons theory. It also holds for the
{\it abelian} (or, rather, $U(1)$) theory, for which there is no quantization in the commutative
case. In the commutative limit the corresponding topologically nontrivial gauge transformations
become singular and decouple from the theory, thus eliminating the need for quantization.
This result will be relevant in the upcoming considerations of the quantum Hall effect.

\section{Connection with fluid mechanics}

At this point we take a break from noncommutative gauge theory to bring into the picture fluid
mechanics and review its two main formulations, Euler and Lagrange. As will become apparent,
the two subjects are intimately related. Already we saw that \nc gauge theory can be formulated
in terms of covariant deformed coordinate operators $X^\mu$. These parallel the spatial
coordinates of particle fluids, with the undeformed background coordinates $x^\mu$ playing
the role of body-fixed labels of the particles. This observation will for the basis for the
formulation of noncommutative fluids. We note that fluids including noncommuting variables
for the description of spin densities have already been studied \cite{JPsf}. In the following
we shall render the whole fluid `stuff' \nc.

\subsection{Lagrange and Euler descriptions of fluids}

We start with a summary review of the two main formulations of fluid mechanics, the particle-fixed
(Lagrange) and space-fixed (Euler) descriptions. For more extensive reviews see \cite{JNP,AK}.

A fluid can be viewed as a dense collection of (identical) particles moving in some $d$-dimensional
space, evolving in time $t$.
The Lagrange description uses the coordinates of the particles comprising the fluid: $X^i (x,t)$.
These are labeled by a set of parameters $x^i$, which are the
coordinates of some fiducial reference configuration and are called
particle-fixed or comoving coordinates. They serve, effectively, as particle `labels'.
Summation over particles amounts to integration over the comoving coordinates $x$ times the
density of particles in the fiducial configuration $\rho_0 (x)$, which is usually taken
to be homogeneous.

In the Euler description the fluid is described by the space-time--dependent density $\rho (r,t)$
and velocity fields $v^i (r,t)$ at each point of space with coordinates $r^i$. 
The two formulations are related by considering the particles at space coordinates $r^i$, that is,
$X^i = r^i$, and
expressing the density and velocity field in terms of the Lagrange variables. We assume
sufficient regularity so that (single-valued) inverse functions $\chi^i (r,t)$ exist:
\be
X^i (t,x) \Bigr|_{x = \chi (t,r)}\!\! =  r^i
\ee
$X^i (x,t)$ provides a mapping of the fiducial particle position $x^i$ to position at time~$t$,
while $\chi^i (r,t)$ is the inverse mapping. The Euler density then is defined by
\be
\rho (r,t) =  \rho_0 \int dx \delta \bigl(X (x,t) - r \bigr)\ .
\label{rhodelta}
\ee
(The integral and the $\delta$-function carry the dimensionality of the relevant
space.) This evaluates as 
\be
\frac1{\rho (r,t)} = \frac{1}{\rho_0} \det \frac{\partial X^i (x,t)}{\partial x^j}\Bigr|_{x =
\chi (r,t)} 
\ee
which is simply the change of volume element from fiducial to real space.
The Euler velocity is 
\be
v^i (r,t) =  \dot{X^i} (x,t) \Bigr|_{x = \chi (r,t)}  
\ee
where overdot denotes differentiation with respect to the explicit time dependence.
(Evaluating an expression  at $x = \chi (r,t)$ is equivalent to eliminating $x$ in favor of
$X$, which is then renamed~$r$.)

The number of particles in the fluid is conserved. This is a trivial (kinematical) condition
in the Lagrange formulation, where comoving coordinates directly relate to particles. In the
Euler formulations this manifests through conservation of the particle current $j^i =\rho v^i$,
given in terms of Lagrange variables by 
\be
j^i (r,t) = \rho_0 \int dx \dot{X^i} (r,t) \delta\bigl( X (x,t) - r \bigr) 
\ee
As a consequence of the above definition it obeys the continuity equation
\be
\dot\rho + \partial_i j^i = 0\ .
\ee

The kinetic part of the lagrangian $K$ for the Lagrange variables is simply the single-particle
lagrangian for each particle in terms of the particle coordinates, $K\sp (X)$, 
summed over all particles.
\be
K =  \rho_0 \int dx K\sp \bigl( X (x,t) \bigr) .
\ee
The exact form of $K\sp$ depends on whether the particles are relativistic or non-relativistic,
the presence of magnetic fields etc. As an example, the kinetic term for a non-relativistic
plasma in an external magnetic field generated by an electromagnetic vector potential ${\cal A}_i$ is
\be
K =  \rho_0 \int dx \left[ \half m \, g_{ij} (X) \, {\dot X}^i {\dot X}^j 
+ q {\cal A}_i (X,t) \, {\dot X}^i \right]
\ee
\label{Kmq}
with $m$ and $q$ the mass and charge of each fluid particle and $g_{ij}$ the metric of space.

Single-particle (external) potentials can be written in a
similar way, while many-body and near-neighbor (density dependent) potentials will be more involved.

\subsection{Reparametrization symmetry and its \nc avatar}

The Lagrange description has an obvious underlying symmetry. Comoving coordinates are essentially
arbitrary particle labels. All fluid quantities are invariant under particle relabeling, that is,
under reparametrizations of the variables $x^i$, provided that the density of the fiducial
configuration $\rho_0$ remains invariant. Such transformations are volume-preserving diffeomorphisms
of the variables $x^i$.

For the minimal nontrivial case of two spatial dimensions, this symmetry corresponds to
area-preserving diffeomorphisms. They can be thought of as canonical transformations on a
two-dimensional phase space and are parametrized by a function of the two spatial variables,
the generator of canonical transformation. Infinitesimal transformations are written
\be
\delta x^i = \epsilon^{ij} \frac{\partial f}{\partial x^j}
\ee
with $f(x)$ the generating function. Obviously $\delta x^i$ satisfies the area-preserving
condition
\be
\det \frac{\partial (x^i + \delta x^i )}{\partial x^j} = 1 ~~~ {\rm or}~~~
\frac{\partial \delta x^i}{\partial x^i} =0
\ee
The same condition can be written in an even more suggestive way. Define a
canonical structure for the two-dimensional space in terms of the Poisson brackets
\be
\{ x^1 , x^2 \} = \theta ~~~ {\rm or}~~~ \{ x^i , x^j \} = \theta \epsilon^{ij} = \theta^{ij}
\ee
for some constant $\theta$. Rescaling $f$ by a factor $\theta^{-1}$, we can re-write $\delta x^i$
as
\be
\delta x^i = \theta^{ij} \partial_j f = \{ x^i , f \}
\ee
Similarly, the transformation of the fundamental (Lagrange) fluid variables under the above
redefinition is
\be
\delta X^i = \partial_j X^i \delta x^j = \theta^{jk} \partial_j X^i \partial_k f
= \{ X^i , f \}
\label{dXflu}
\ee

The above look like the classical analog (or precursor) of the
gauge transformations of the covariant noncommutative gauge coordinates $X^i$ of the previous
sections. This is not accidental: the area-preserving transformations for the fluid correspond
to relabeling the parameters $x$ and do not generate a physically distinct fluid
configurations. They represent simply a redundancy in the description of the fluid in terms
of Lagrange coordinates; that is, a gauge symmetry. Physical fluid quantities, such as the
Euler variables, or the fluid lagrangian, are expressed as integrals of quantities transforming
`covariantly' under the above transformation; that is, transforming by the Poisson bracket of the
quantity with the generator of the transformation $f$, as in (\ref{dXflu}). They are, therefore,
invariant under such transformations; that is, gauge invariant.

The analogy with \nc gauge theory becomes manifest by writing the Lagrange particle coordinates
in terms of their deviation from the fiducial coordinates \cite{BaSu}-\cite{JP}:
\be
X^i (x,t) = x^i + a^i (x,t) = x^i + \theta^{ij} A_j (x,t)
\label{XaA}
\ee
The deviation $a^i$, and its dual $A_i$ do not transform covariantly any more; rather
\be
\delta A_i = \partial_i f + \{ A_i , f \}
\ee
The similarity with the gauge transformation of a gauge field is obvious. The duals of the $X^i$
\be
D_i = \omega_{ij} X^j = \omega_{ij} x^j + A_i
\ee
obviously correspond to covariant derivatives (although at this stage they are just rewritings
of the comoving particle coordinates). The analog of the field strength is
\be
{\hat F}_{ij} = \{ D_i , D_j \} = \omega_{ij} + \partial_i A_j - \partial_j A_i + \{ A_i , A_j \}
\ee
This is related to the fluid density, which in the Poisson bracket formulation reads
\be 
\frac{\rho_0}{\rho} = \det \frac{\partial X^k (x,t)}{\partial x^l} = \frac{1}{\theta} \{ X^1 , X^2 \}
\label{rhoXX}
\ee
The field strength calculates as:
\be
{\hat F}_{ij} = \omega^{ij} \{ X^1 , X^2 \} = \frac{\rho_0}{\rho} \epsilon_{ij}
\label{Frho}
\ee
The field strength essentially becomes the (inverse) fluid density!

Similar considerations generalize to higher dimensions, with one twist: canonical transformations,
the classical version of \nc gauge transformations,
are only a symplectic subgroup of full volume-preserving diffeomorphisms. Higher-dimensional \nc
gauge theory is analogous to a special version of fluid mechanics that enjoys a somewhat limited
particle relabeling invariance. For the purposes of describing the quantum Hall effect, an
essentially two-dimensional situation, this is inconsequential.

\subsection{Gauging the symmetry}

In the above discussion the role of time was not considered. The particle relabeling ($x$-space 
reparametrization) considered above were time-independent. Time-dependent transformations are not,
a priori, invariances of the fluid since they introduce extra, nonphysical terms in the
particle velocities ${\dot X}^i (x,t)$. To promote this transformation into a full space-time
gauge symmetry we must gauge time derivatives by introducing a temporal gauge field $A_0$:
\be
D_0 X^i = {\dot x}^i + \{ A_0 , X^i \}
\ee
Under the transformation (\ref{dXflu}) with a time-dependent function $f$ the above derivative
will transform covariantly
\be
\delta D_0 X^i = \{ D_0 X^i , f \}
\ee
provided that the gauge field $A_0$ transforms as
\be
\delta A_0 = {\dot f} + \{ A_0 , f \}
\ee
This gauging, however, has dynamical consequences. We can gauge fix the theory by choosing the
temporal gauge, putting $A_0 =0$. The action becomes identical to the ungauged action, with the
exception that now we have to satisfy the Gauss law for the gauge-fixed symmetry, that is, the
equation of motion for the reduced field $A_0$. The exact form of the constraint depends on the
kinetic term of the lagrangian for the fluid:
\be
G = \{ X^i , \frac{\partial K}{\partial {\dot X}^i} \} = 0
\ee
As an example, for the plasma of (\ref{Kmq})
the Gauss law reads
\be
G = \{ {\dot X}^i , m g_{ij} (X) \, {\dot X}^j + q {\cal A}_i (X) \} = 0
\ee
Interesting two-dimensional special cases are ($g_{ij} = \delta_{ij}$, $q=0$), when
\be
G = \{ {\dot X}^i , X^i \} = 0
\ee
and the `lowest Landau level' case of massless particles in a constant magnetic field
($m=0$, ${\cal A}_i = (B/2) \epsilon_{ij} X^j$), when
\be
G = \{ X^1 , X^2 \} = 0
\ee

We conclude by mentioning that the fluid structure we described in this section can also
be interpreted as membrane dynamics. Indeed, a membrane is, in principle, a sheet of fluid
in a higher-dimensional space. A two-dimensional membrane in two space dimensions is
space-filling, and thus indistinguishable from a fluid, the density expressing the
way in which the membrane shrinks or expand locally. The full correspondence of membranes,
noncommutative (matrix) theory and fluids, relativistic and non-relativistic, has been
examined elsewhere \cite{MembMatFlu}. We shall not expand on it here.

\subsection{\Nc fluids and the Seiberg-Witten map}

In the previous section we alluded to the connection between \nc gauge theory and
fluid mechanics. It is time to make the connection explicit \cite{JP}. We shall work specifically
in two (flat) spatial dimensions, as the most straightforward case and relevant to the
quantum Hall effect.

The transition from (classical) fluids to \nc fluids is achieved the same way
as the transition from classical to quantum mechanics. We promote the canonical 
Poisson brackets introduced in the previous section to (operator) commutators.
All Poisson brackets that appear become commutators:
\be
\{ ~ , ~ \} ~ \to ~ -i [ ~ , ~ ]
\ee
So the comoving parameters satisfy
\be
[ x^i , x^j ] = i\theta^{ij}
\ee
They have become a noncommutative plane. This means that the particle labels cannot have
`sharp' values and pinpointing the particles of the fluid is no more possible. In effect,
we have a `fuzzification' of the underlying fluid particles and a corresponding `fuzzy' fluid.

The remaining structure smoothly goes over to \nc gauge theory, as already alluded.
We assume that the noncommutative coordinates $x^1$, $x^2$ act on a single irreducible
representation of their Heisenberg algebra; this effectively assigns a single particle state
for each `point' of space (each state in the representation). Inclusion of multiple copies of the
irreducible representations would correspond to multiple particle states per `point' of space and
would endow the particles with internal degrees of freedom.

Integration over the comoving parameters becomes $2\pi \theta$ times trace over the
representation space. Summation over particles, then, becomes
\be
\sum_{\rm particles} = \rho_0 \int dx ~ \to ~ 2\pi \theta \rho_0 \Tr
\ee
The parameter $\theta$, or its inverse $\omega$, was introduced arbitrarily and plays no role
in the fluid description. This is similar to the background-independent formulation of \nc
gauge theory in terms of covariant derivatives or coordinates. Presently, we
relate $\theta$ to the inverse density of the fiducial configuration $\rho_0^{-1}$
\be
2\pi\theta = \frac{1}{\rho_0}
\ee
in which case the factor in the preceding equation disappears. Particle summation becomes
a simple trace, so particles are identified with states in the representation space. This
relation between fiducial density and noncommutativity parameter will always be
assumed to hold from now on.

The Lagrange coordinates of particles $X^i$ and the gauge field $A_0$ are functions of the
underlying `fuzzy' (\nc) particle
labels, and thus become \nc fields. Area-preserving reparametrizations, which are canonical
transformations in the classical case, become unitary transformations in the \nc case (think,
again, of quantum mechanics). Operators $X^i$ transform by unitary conjugations; infinitesimally,
\be
\delta X^i = i [f, X^i ]
\ee
The deviations of $X^i$ from the fiducial coordinates $x^i$, on the other hand, as defined
in (\ref{XaA}), and the temporal gauge field pick up extra terms and transform as proper gauge
fields:
\be
\delta A_\mu = \partial_\mu f -i [ A_\mu , f ]
\ee

The remaining question is the form of the (gauge invariant) lagrangian that corresponds to the
\nc fluid. This depends on the specific fluid dynamics and will be dealt with in the next section.
Before we go there, we would like to examine further the properties of the \nc fluid that derives
from the present construction. Just because the underlying particles become fuzzy does not
necessarily mean that the emerging fluid cannot be described in traditional terms. Indeed, fluids
are dense distributions of particles and we are not supposed to be able to distinguish individual
particles in any case. The Euler description, which talks about collective fluid properties like
density and velocity, remains valid in the \nc case as we shall see.

The \nc version of equation (\ref{rhoXX}) for the density becomes (with $2\pi\theta\rho_0 =1$)
\be 
[ X^1 , X^2 ] = \frac{i}{2\pi\rho}
\label{ncrhoXX}
\ee
This relation would suggest
that the density, too, becomes a \nc field. The difficulty with this expression is that it gives
the density as a function of the underlying comoving coordinates, which we know are noncommutative.

A better expression is (\ref{rhodelta}), which gives the density as a function of a point in space $r$.
This formula directly transcribes into
\be
\rho(r,t) = \Tr \delta \bigl(X - r \bigr)
\ee
in the \nc case. $r$ is still an ordinary space variable, and the trace eliminates the operator
nature of the expression in the right hand side, rendering a classical function of $r$ and $t$.
The only difficulty is in the definition of the delta function for the noncommutative argument
$X^i - r^i$: the various $X^i$ (two in our case) are operators and do not commute, so there are
ordering issues in defining any function of the two. In fact, the operator $\delta (X-r)$ may not
even be hermitian unless properly ordered, which would produce a complex density.

In dealing with such problems, a procedure similar to the definition of the `symbol' of a \nc field
is followed: a standard ordering of all monomials involving various $X^i$s is prescribed. The Weyl
(totally symmetrized) ordering is usually adopted. Under this ordering, the delta function above
is defined as
\be
\delta \bigl(X - r \bigr) = \int dk e^{i k_i (r^i - X^i )}
\ee
where $k_i$ are classical (c-number) Fourier integration parameters. The above operator has also
the advantage of being hermitian. The spatial Fourier transform of the density with respect to 
$r$ is simply
\be
\rho(k,t) = \Tr \, e^{-i k_i X^i}
\label{rhok}
\ee
In a similar vein, we use the classical expression for the 
particle current
\be
j^i (r,t) = \rho_0 \int dx {\dot X}^i \delta \bigl(X - r \bigr)
\ee
to write the corresponding expression for the \nc fluid as
\be
j^i (k,t) = \Tr \, D_0 X^i \, e^{-i k_j X^j}
\label{jk}
\ee
In the above, we used the covariant time
derivative in order to make the expression explicitly gauge invariant. The corresponding
current is real, as the trace ensures that the change of ordering between $D_0 X$ and the exponential
is immaterial.

The crucial observation is that the above density and current still satisfy the
continuity equation, which in Fourier space becomes
\be
{\dot \rho} + i k_i j^i = 0
\ee
The proof is straightforward and relies on the following two facts, true due to the cyclicity of trace:
\be
\frac{d}{dt} \Tr \, e^{-i k_i X^i} = -i \Tr \, k_j {\dot X}^j e^{-i k_i X^i}
\ee
and
\be
\Tr \, [A_0 , k_j X^j ] e^{-i k_i X^i} = 0
\ee
The \nc fluid, therefore, has an Euler description in terms of a traditional conserved particle 
density and current.

The above observation is the basis for a mapping between commutative and \nc gauge theories,
which fist arose in the context of string theory and is known as the Seiberg-Witten map \cite{SW}.
The key element is that, in 2+1 dimensions, a conserved current can be written in terms
of its dual two-form, which then satisfies the Bianchi identity.
Specifically, define
\be
J_{\mu \nu} = \epsilon_{\mu \nu \lambda} j^\lambda
\ee
where $j^0 = \rho$. Then, due to the continuity equation $\partial_\mu j^\mu =0$,
$J_{\mu \nu}$ satisfies
\be
\partial_\mu J_{\nu \lambda} + {\rm cyclic~perms.} = 0 ~~~ {\rm or} ~~~ \dd {\rm J}=0
\ee
This means that $\rm J$ can be considered as an abelian field strength, which allows us
to define an abelian commutative gauge field ${\tilde A}_\mu$. The
reference configuration of the fluid, in which particles are in their fiducial positions
$X^i = x^i$ and corresponds to vanishing \nc gauge field, gives $j_0^\mu = (\rho_0 , 0,0)$
or ${\rm J}_0 = \rho_0 \dd x^1 \dd x^2$.
If we want to have this configuration correspond to vanishing abelian gauge field
${\tilde F}_{\mu \nu}$, we have to define
\be
{\tilde \F} = {\rm J} - {\rm J}_0
\ee
or, more explicitly
\be
{\tilde F}_{0i} = \epsilon_{ik} j^k ~,~~~
{\tilde F}_{ij} = \epsilon_{ij} ( \rho - \rho_0 )
\ee
Substituting the explicit expressions (\ref{rhok},\ref{jk}) for $\rho$ and $j^i$, and expressing
$X^i$ in them in terms of noncommutative fields, gives an explicit mapping between
the \nc fields $A_\mu$ and the commutative fields ${\tilde A}_\mu$.

Similar considerations extend to higher dimensions but, again, we shall not dwell on them here
\cite{JP}-\cite{APSW}.
The moral lesson of the above is that the Lagrange formulation of fuzzy fluids is inherently
noncommutative, while the Euler formulation is commutative. The Seiberg-Witten map between them
becomes the transition from the particle-fixed Lagrange to the space-fixed Euler formulation.

\section{The \nc description of quantum Hall states}

We reach, now, one of the main topics of this presentation. Is the above useful to anything?
Can we use it to describe or solve any physical system or does it remain an interesting
peculiarity?

To find an appropriate application, we must look for systems with `fuzzy' particles. This is not
hard: quantum mechanical particles on their phase spaces are fuzzy, due to Heisenberg uncertainty.
This can be carried through, and eventually leads to the description of one-dimensional fermions
in terms of matrix models.

A more interesting situation arises in lowest Landau level physics, in which particles become fuzzy
on the {\it coordinate} space. Spatial coordinates become noncommuting when restricted to the lowest
Landau level \cite{APcs,DuJT}, already introducing a \nc element (although quite distinct from the 
one introduced
in the sequel). This is also the setting for the description of quantum Hall states
and will be the topic of the present section.

\subsection{\Nc Chern-Simons description of the quantum Hall fluid}

The system to be described consists of a large number $N \to \infty$ of electrons 
on the plane in the lowest Landau level of an external constant magnetic field $B$
(we take the electron charge $e=1$).
Upon proper dynamical conditions, they form quantum Hall states (for a review of
the quantum Hall effect see \cite{Laug}.)
According to the observations of the previous section, we can parametrize their
coordinates as a fuzzy fluid in terms of two \nc Lagrange coordinates
(infinite hermitian `matrices') $X^i$, $i=1,2$, that is, by two
operators on an infinite Hilbert space. The density of these electrons is not
fixed at this point, but will eventually relate to the noncommutativity parameter as 
$\rho_0 = 1/2\pi \theta$.

The action is the \nc fluid analog of the gauge action of massless particles in an
external constant magnetic field.
In the symmetric gauge for the magnetic field, this would read
\be
S = \int dt\, \frac{B}{2} \, \Tr \left\{ \epsilon_{ij} D_0 X^i \, X^j\right\}
= \int dt\, \frac{B}{2} \, \Tr \left\{ \epsilon_{ij} ({\dot X}^i 
+i [A_0 , X^i ]) X^j \right\}
\ee
The above expression was made gauge invariant by gauging the time derivative and
introducing a \nc temporal gauge field $A_0$. As explained in previous sections, 
however, this introduces a Gauss law constraint, which in the present case reads
\be
[ X^1 , X^2 ] = 0
\ee
This is undesirable in many ways. The would-be \nc coordinates become commutative,
eliminating the fuzziness of the description. More seriously, the density of the
fluid classically becomes singular, as can be seen from the expression (\ref{ncrhoXX})
for the inverse fluid density. (It can also be deduced from the commutative expression
(\ref{rhok}), although in a slightly more convoluted way.)

Taking care of the above difficulty also gives the opportunity to introduce an important
piece of physics for the system: fractional quantum Hall states (Laughlin states,
in their simplest form) are incompressible and have a constant spatial density $\rho_0$.
The {\it filling fraction} $\nu$ of the state is defined as the fraction of the Landau level
density $\rho_{_{LL}} = B/2\pi$ that $\rho_0$ represents:
\be
\nu = \frac{\rho_0}{\rho_{_{LL}}} = \frac{2\pi\rho}{B} = \frac{1}{\theta B}
\ee
where the \nc parameter $\theta$ is related to the desired fluid density in the standard
way, spelled out again as
\be
\rho_0 = \frac{1}{2\pi\theta}
\ee
We can introduce this constant density $\rho_0$ in the system by modifying the Gauss law 
constraint by an appropriate constant, achieved by adding a term linear in $A_0$.
The resulting action reads
\be
S = \int dt\, \frac{B}{2} \, \Tr \left\{ \epsilon_{ij} ({\dot X}^i 
+i [A_0 , X^i ]) X^j + 2\theta A_0 \right\}
\label{CSX}
\ee
This was first proposed by Susskind \cite{Susskind}, 
motivated by the earlier, classical mapping of the quantum Hall fluid to a gauge action \cite{BaSu}
and related string theory work \cite{STqhe}.
The equation of motion for $A_0$, now, imposes the Gauss law constraint
\be
[ X^1 , X^2 ] = i \theta
\label{XX}
\ee
essentially identifying $X^1$,$X^2$ with a \nc plane.

Interestingly, the above action is exactly the noncommutative CS action in 2+1 dimensions!
A simple comparison of expression (\ref{CSDD}) and (\ref{CSX}) above reveals that they are
the same, upon identifying $\theta D_i = \epsilon_{ij} X^j$.
The coefficient of the CS term $\lambda$ relates to $B$ and the filling fraction as
\be
\lambda = \frac{B \theta}{4\pi} = \frac{1}{4\pi \nu}
\label{lambda}
\ee
This establishes the connection of the \nc Chern-Simons action with
the quantum Hall effect.

As before, gauge transformations are conjugations of $X^i$ or $D_i$ by arbitrary
time-dependent unitary operators. In the quantum Hall fluid context they
take the meaning of reshuffling the electrons. Equivalently, the
$X^i$ can be considered as coordinates of a two-dimensional fuzzy
membrane, $2\pi\theta$ playing the role of an area quantum and gauge 
transformations realizing area preserving diffeomorphisms.
The canonical conjugate of $X^1$ is $P_2 = B X^2$, and the 
generator of gauge transformations is
\be
G = -i B [ X^1, X^2 ] = B\theta = \frac{1}{\nu}
\label{Gauss}
\ee
by virtue of (\ref{XX}). Since gauge transformations are interpreted
as reshufflings of particles, the above has the interpretation of 
endowing the particles with quantum statistics of order $1/\nu$.

\subsection{Quasiparticle and quasihole classical states}

The classical equation (\ref{XX}) has a unique solution, modulo
gauge (unitary) transformations, namely the unique irreducible
representation of the Heisenberg algebra. Representation states can be
conveniently written in a Fock basis $|n\ra$, $n=0,1,\dots$,
for the ladder operators $X^1 \pm i X^2$, $|0\ra$ representing a 
state of minimal spread at the origin. 
The classical theory has this representation as its unique state, 
the vacuum.

Deviations from the vacuum (\ref{XX}) can be achieved by introducing
sources in the action \cite{Susskind}. A localized source at the origin 
has a density
of the form $\rho = \rho_0 - q \delta^2 (x)$ in the continuous
(commutative) case, representing a point source of particle number
$-q$, that is, a hole of charge $q$ for $q>0$. The noncommutative 
analog of such a density is
\be
[ X^1 , X^2 ] = i \theta ( 1 + q | 0 \ra \la 0 | )
\label{q}
\ee
In the membrane picture the right-hand side
of (\ref{q}) corresponds to area and implies that the area quantum
at the origin has been increased to $2\pi\theta(1+q)$, therefore
piercing a hole of area $A = 2\pi\theta q$ and creating a particle 
deficit $q = \rho_0 A$. We shall call this a quasihole state.
For $q>0$ we find the quasihole solution of (\ref{q}) as
\be
X^1 + i X^2 = \sqrt{2\theta} \sum_{n=1}^\infty \sqrt{n+q}
\, | n-1 \ra \la n |
\label{Xqpos}
\ee
Such solutions are called \nc gauge solitons \cite{APA,GN,GMS,JaMaWa,HKL}.

The case of quasiparticles, $q<0$ is more interesting. 
Clearly the area quantum cannot be diminished below zero, and 
equations (\ref{q}) and (\ref{Xqpos}) cannot hold for $-q>1$. 
The correct equation is, instead,
\be
[ X^1 , X^2 ] = i \theta \left( 1 - \sum_{n=0}^{k-1} 
| n \ra \la n | - \epsilon | k \ra \la k | \right)
\label{qneg}
\ee
where $k$ and $\epsilon$ are the integer and fractional part
of the quasiparticle charge $-q$. The solution of (\ref{qneg}) is
\be
X^1 + i X^2 = \sum_{n=0}^{k-1} z_n | n \ra \la n | + 
\sqrt{2\theta} \sum_{n=k+1}^\infty \sqrt{n-k-\epsilon} 
\, | n-1 \ra \la n |
\label{Xneg}
\ee
(For $k=0$ the first sum in (\ref{qneg},\ref{Xneg}) drops.) 
In the membrane picture, $k$ quanta of the membrane have `peeled' 
and occupy positions $z_n = x_n + i y_n$ on the plane, while the
rest of the membrane has a deficit of area at the origin equal
to $2\pi \theta \epsilon$, leading to a charge surplus $\epsilon$.
Clearly the quanta are electrons that sit on top of the continuous
charge distribution. If we want all charge density to be concentrated
at the origin, we must choose all $z_n =0$.
The above quasiparticle states for integer $q$ are
the noncommutative solitons and flux tubes that are also
solutions of noncommutative gauge theory, while the quasihole states
are not solutions of the \nc gauge theory action and have no direct analog.

Laughlin theory predicts that quasihole excitations in the 
quantum Hall state have their charge $-q$ quantized in integer units
of $\nu$, $q = m \nu$, with $m$ a positive integer. We see that 
the above discussion gives no hint of this quantization, while 
we see at least some indication of electron quantization in 
(\ref{qneg},\ref{Xneg}). Quasihole quantization will emerge in 
the quantum theory, as we shall see shortly, and is equivalent to
a quantization condition of the \nc Chern-SImons term.

\subsection{Finite number of electrons: the Chern-Simons matrix model}

Describing an infinitely plane filled with electrons is not the most
interesting situation. We wish to describe quantum Hall states of
finite extent consisting of $N$ electrons. Obviously the coordinates 
$X^i$ of the \nc fluid description would have to be represented by 
finite $N \times N$ matrices.
The action (\ref{CSX}), however, and the equation (\ref{XX}) to
which it leads, are inconsistent for finite matrices, and a modified
action must be written which still captures the physical features
of the quantum Hall system. Such an action exists, and leads to a
matrix model truncation of the \nc Chern-Simons action involving
a `boundary field' \cite{APqh}. It is
\be
S = \int dt \frac{B}{2} \Tr \left\{ \epsilon_{ij} ({\dot X}^i 
+i [A_0 , X^i ]) X^j + 2\theta A_0 - \omega (X^i)^2 \right\} 
+ \Psi^\dagger (i {\dot \Psi} - A_0 \Psi)
\label{CSPsi}
\ee
It has the same form as the planar CS action, but with two extra
terms. The first, and most crucial, involves $\Psi$, a complex $N$-vector that 
transforms in the fundamental of the gauge group $U(N)$: 
\be
X^i \to U X^i U^{-1} ~,~~~\Psi \to U \Psi
\ee
Its action is a covariant kinetic term similar to a complex scalar
fermion. We shall, however, quantize it as a boson; this is
perfectly consistent, since there is no spatial kinetic term that
would lead to a negative Dirac sea and the usual inconsistencies
of first-order bosonic actions.

The term proportional to $\omega$ (not to be confused with $\theta^{-1}$)
serves as a spatial regulator:
since we will be describing a finite number of electrons, there 
is nothing to keep them localized anywhere in the plane. 
We added a confining harmonic potential which serves as a `box' 
to keep the particles near the origin.

We can again impose the $A_0$ equation of motion as a Gauss
constraint and then put $A_0 =0$. In our case it reads
\be
G \equiv -iB [ X^1 , X^2 ] + \Psi \Psi^\dagger - B\theta =0
\label{G}
\ee
Taking the trace of the above equation gives
\be
\Psi^\dagger \Psi = NB \theta
\ee
The equation of motion for $\Psi$ in the $A_0 =0$ gauge is 
${\dot \Psi} =0$. So we can take it to be
\be
\Psi = \sqrt{NB\theta} \, |v\ra
\ee
where $|v\ra$ is a constant vector of unit length. Then
(\ref{G}) reads
\be
[ X^1 , X^2 ] = i\theta \left( 1 - N |v\ra \la v| \right)
\label{XXv}
\ee
This is similar to (\ref{XX}) for the infinite plane case,
with an extra projection operator. Using the residual gauge freedom
under time-independent unitary transformations, we can rotate
$|v\ra$ to the form $|v\ra = (0,\dots 0,1)$. The above commutator
then takes the form $i\theta \,{diag}\, (1,\dots, 1, 1-N)$ which
is the `minimal' deformation of the planar result (\ref{XX})
that has a vanishing trace. 

In the fluid (or membrane) picture, $\Psi$
is like a boundary term. Its role is to absorb the `anomaly' of the
commutator $[X^1 , X^2 ]$, much like the case of a boundary
field theory required to absorb the anomaly of a bulk (commutative)
Chern-Simons field theory.

The equations of motion for $X^i$ read
\be
{\dot X}^i + \omega \epsilon_{ij} X^j =0
\ee
This is just a matrix harmonic oscillator. It is solved by
\be
X^1 + i X^2 = e^{i\omega t} A
\label{Arot}\ee
where $A$ is any $N \times N$ matrix satisfying the constraint
\be
[ A, A^\dagger ] = 2\theta (1 - N |v\ra \la v| )
\label{AA}
\ee

The classical states of this theory are given by the set of
matrices $A = X^1 + i X^2$ satisfying (\ref{AA}) or (\ref{XXv}).
We can easily find them by choosing a basis in which one of
the $X$s is diagonal, say, $X^1$. Then the commutator 
$[ X^1 , X^2 ]$ is purely off-diagonal and the components of
the vector $|v\ra$ must satisfy $|v_n |^2 = 1/N$. We can use
the residual $U(1)^N$ gauge freedom to choose the 
phases of $v_n$ so that $v_n = 1/\sqrt{N}$. So we get
\be
(X^1 )_{mn} = x_n \delta_{mn} ~,~~~
(X^2 )_{mn} = y_n \delta_{mn} + \frac{i\theta}
{x_m - x_n} (1-\delta_{mn} )
\label{Xsol}
\ee
The solution is parametrized by the $N$ eigenvalues of $X^1$, 
$x_n$, and the $N$ diagonal elements of $X^2$, $y_n$. 

\subsection{Quantum Hall `droplet' vacuum}

Not all solutions found above correspond to quantum Hall fluids. 
In fact, choosing all $x_n$ and $y_n$ much bigger than $\sqrt\theta$
and not too close to each other, both $X^1$ and $X^2$ become
almost diagonal; they represent $N$ electrons scattered in 
positions $(x_n , y_n )$ on the plane and performing rotational
motion around the origin with angular velocity $\omega$. This
is the familiar motion of charged particles in a magnetic field 
along lines of equal potential when their proper kinetic term
is negligible. Quantum Hall states will form when particles
coalesce near the origin, that is, for states of low energy. 

To find the ground state, we must minimize the potential
\be
V = \frac{B\omega}{2} \Tr [(X^1)^2 + (X^2)^2 ] = 
\frac{B\omega}{2} \Tr (A^\dagger A)
\ee
while imposing the constraint (\ref{XXv}) or (\ref{AA}). This
can be implemented with a matrix Lagrange multiplier $\Lambda$
(essentially, solving the equations of motion including $A_0
\equiv \Lambda$
and putting the time derivatives to zero). We obtain
\be
A = [\Lambda , A] ~,~~~~{\rm or}~~~ X^i = i \epsilon_{ij}
[ \Lambda , X^j ]
\label{AAL}
\ee
This is reminiscent of canonical commutation relations for a
quantum harmonic oscillator, with $\Lambda$ playing the role 
of the hamiltonian. We are led to the solution
\be
A = \sqrt{2\theta} \sum_{n=0}^{N-1} \sqrt{n} |n-1\ra \la n | 
~,~~~  \Lambda = \sum_{n=0}^{N-1} n |n\ra \la n | 
~,~~~ |v\ra = | N-1 \ra
\label{APN}
\ee
This is essentially a quantum harmonic oscillator and 
hamiltonian projected to the lowest $N$ energy eigenstates. 
It is easy to check that the above satisfies both 
(\ref{AA}) and (\ref{AAL}). 
Its physical interpretation is clear: 
it represents a circular quantum Hall `droplet' of radius 
~$\sqrt{2N\theta}$. Indeed, the radius-squared matrix 
coordinate $R^2$ is
\beqar
R^2 &=& (X^1)^2 + (X^2)^2 = A^\dagger A + \half [A,A^\dagger] \\
    &=& \sum_{n=0}^{N-2} \theta (2n+1) |n\ra \la n| +
\theta (N-1) |N-1 \ra \la N-1 |
\eeqar
The highest eigenvalue of $R^2$ is $(2N-1)\theta$. The particle 
density of this droplet is $\rho_0 = N/(\pi R^2) \sim 
1/(2\pi \theta)$ as in the infinite plane case. 

The matrices $X^i$ are known and can be explicitly diagonalized 
in this case. Their eigenvalues are given by the
zeros of the $N$-th Hermite polynomial (times $\sqrt{2\theta}$).
In the large-$N$ limit the distribution of these zeros obeys the 
famous Wigner semi-circle law, with radius $\sqrt N$. Since these
eigenvalues are interpreted as electron coordinates, this
confirms once more the fact that the electrons are evenly 
distributed on a disk of radius $~\sqrt{2N\theta}$.

\subsection{Excited states of the model}

Excitations of the classical ground state can now be considered.
Any perturbation of (\ref{APN}) in the form of (\ref{Xsol}) is,
of course, some excited state. We shall concentrate, however,
on two special types of excitations. 

The first is obtained by performing on $A,A^\dagger$ all
transformations generated by the infinitesimal transformation
\be
A' = A + \sum_{n=0}^{N-1} \epsilon_n (A^\dagger)^n
\label{dropex}
\ee
with $\epsilon_n$ infinitesimal complex parameters. The sum is
truncated to $N-1$ since $A^\dagger$ is an $N \times N$ matrix
and only its first $N$ powers are independent. It is
obvious that (\ref{AA}) remains invariant under the above
transformation and therefore also under the finite transformations
generated by repeated application of (\ref{dropex}). 

If $A,A^\dagger$
were true oscillator operators, these would be canonical (unitary)
transformations, that is, gauge transformations that would
leave the physical state invariant. For the finite $A,A^\dagger$
in (\ref{APN}), however, these are {\it not} unitary transformations
and generate a new state. To understand what is that new state,
examine what happens to the `border' of the circular quantum Hall
droplet under this transformation. This is defined by 
$A^\dagger A \sim 2N\theta$ (for large $N$). To find the new boundary
parametrize $A \sim \sqrt{2N\theta} e^{i\phi}$, with $\phi$ the 
polar angle on the plane and calculate $(A^\dagger A)'$. 
The new boundary in polar coordinates is
\be
R' (\phi ) = \sqrt{2N\theta} + \sum_{n=-N}^N c_n e^{in\phi} 
\label{newR}
\ee
where the coefficients $c_n$ are
\be
c_n = c_{-n}^* = \frac{R^n}{2} \epsilon_{n-1}
~~(n>0),~~~~c_0 =0
\label{cc}
\ee
This is an arbitrary area-preserving deformation of the
boundary of the droplet, truncated to the lowest $N$ 
Fourier modes. The above states are, therefore, 
arbitrary area-preserving boundary excitations of the
droplet \cite{Wen,IKS,CTZ}, appropriately truncated to reflect
the finite noncommutative nature of the system (the fact
that there are only $N$ electrons). 

Note that on the plane there is an infinity of 
area-preserving diffeomorphisms that produce a specific 
deformation of a given curve. From the droplet point of view,
however, these are all gauge equivalent since they deform
the outside of the droplet (which is empty) or the inside
of it (which is full and thus invariant). The finite theory
that we examine has actually broken this infinite gauge
freedom, since most of these canonical transformations of
$a,a^\dagger$ do not preserve the Gauss constraint (\ref{AA})
when applied on $A,A^\dagger$. The transformations (\ref{dropex})
pick a representative in this class which respects the constraint.

The second class of excitations are the analogs of quasihole and
quasiparticle states. States with a quasihole of charge $-q$ at 
the origin can be written quite explicitly in the form
\be
A = \sqrt{2\theta} \left( \sqrt{q} |N-1 \ra \la 0| + \sum_{n=1}^{N-1} 
\sqrt{n+q} |n-1\ra \la n| \right) ~,~~~ q>0
\label{qhole}
\ee
It can be verified that the eigenvalues of $A^\dagger A$ are
\be
(A^\dagger A)_n = 2\theta(n+q) ~,~~~ n=0,1,\dots N-1
\label{eigen}
\ee
so it represents a circular
droplet with a circular hole of area $2\pi\theta q$ at the origin,
that is, with a charge deficit $q$. The droplet radius has 
appropriately swelled, since the total number of particles is
always $N$. 

Note that (\ref{qhole}) stills respects the Gauss
constraint (\ref{AA}) (with $|v\ra = |N-1\ra$) {\it without} the
explicit introduction of any source. So, unlike the infinite
plane case, this model contains states representing quasiholes
without the need to introduce external sources. What happens
is that the hole and the boundary of the droplet together cancel
the anomaly of the commutator, the outer boundary part absorbing
an amount $N+q$ and the inner (hole) boundary producing
an amount $q$. This possibility did not exist in the infinite
plane, where the boundary at infinity was invisible,
and an explicit source was needed to nucleate the hole.

Quasiparticle states are a different matter. In fact, there
are {\it no} quasiparticle states with the extra particle number
localized anywhere within the droplet. Such states do not belong
to the $\nu = 1/B\theta$ Laughlin state. There are quasiparticle
states with an {\it integer} particle number $-q=m$, and the
extra $m$ electrons occupying positions {\it outside} the droplet.
The explicit form of these states is not so easy to write.
At any rate, it is interesting that the matrix model `sees' the
quantization of the particle number and the inaccessibility of
the interior of the quantum Hall state in a natural way.

Having said all that, we are now making the point that {\it
all types of states defined above are the same}. Quasihole
and quasiparticle states are nonperturbative boundary excitations
of the droplet, while perturbative boundary excitations can
be viewed as marginal particle states. 

To clarify this point,
note that the transformation (\ref{dropex}) or (\ref{newR})
defining infinitesimal boundary excitations has $2N$ real
parameters. The general state of the system, as presented in
(\ref{Xsol}) also depends on $2N$ parameters (the $x_n$ and $y_n$).
The configuration space is connected, so all states can be reached
continuously from the ground state. Therefore, all states can be
generated by exponentiating (\ref{dropex}). This is again a feature
of the finite-$N$ model: there is no sharp distinction between
`perturbative' (boundary) and `soliton' (quasiparticle) states, 
each being a particular limit of the other.

\subsection{Equivalence to the Calogero model}

The model examined above should feel very familiar to Calogero model aficionados.
Indeed, it is equivalent to the harmonic rational
Calogero model \cite{Calogero,Sutherland,Moser}, whose
connection to fractional statistics \cite{PolFS} and anyons \cite{BHKV}-\cite{Ouv}
has been established in different contexts. This is
an integrable system of $N$  nonrelativistic particles on the line 
interacting with mutual inverse-square
potential and an external harmonic potential, with hamiltonian
\be
H = \sum_{n=1}^N \left( \frac{\omega}{2B} p_n^2 + \frac{B\omega}{2}
x_n^2 \right) + \sum_{n\neq m} \frac{\nu^{-2}}{(x_n - x_m )^2}
\label{Hcalo}
\ee 
In terms of the parameters of the model, the mass of the particles
is $B/\omega$ and the coupling constant of the two-body inverse-square
potential is $\nu^{-2}$. We refer the reader to \cite{OP,PolMM,Polrev}
for details on the Calogero model and its connection with the matrix model. 
Here we simply state the relevant
results and give their connection to quantum Hall quantities.

The positions of the Calogero particles $x_n$ are the
eigenvalues of $X^1$, while the momenta $p_n$ are the diagonal
elements of $X^2$, specifically $p_n = B y_n$. The motion of the
$x_n$ generated by the hamiltonian (\ref{Hcalo}) is compatible with
the evolution of the eigenvalues of $X^1$ as it evolves in time
according to (\ref{Arot}). So the Calogero model gives a 
one-dimensional perspective of the quantum Hall state by monitoring
some effective electron coordinates along $X^1$ (the eigenvalues
of $X^1$).

The hamiltonian of the Calogero model (\ref{Hcalo}) is equal
to the the matrix model potential $V = \half B\omega \Tr (X^i)^2$.
Therefore, energy states map between the two models.
The ground state is obtained by putting the particles at their
static equilibrium positions. Because of their repulsion, they
will form a lattice of points lying at the roots of the $N$-th 
Hermite polynomial and reproducing the semi-circle Wigner 
distribution mentioned before. 

Boundary excitations of the quantum Hall
droplet correspond to small vibrations around the equilibrium
position, that is, sound waves on the lattice. Quasiholes are
large-amplitude (nonlinear) oscillations of the particles at a
localized region of the lattice. For a quasihole of charge $q$
at the center, on the average $q$ particles near $x=0$ participate
in the oscillation. 

Finally, quasiparticles are excitations
where one of the particles is isolated outside the ground state
distribution (a `soliton') \cite{PolWS}. 
As it moves, it `hits' the distribution
on one side and causes a solitary wave of net charge 1 to 
propagate through the distribution. As the wave reaches the 
other end of the distribution another particle emerges and
gets emitted there, continuing its motion outside the distribution.
So a quasiparticle is more or less identified with a Calogero
particle, although its role, at different times, is assumed by
different Calogero particles, or even by soliton waves within the
ground state distribution.

Overall, we have a `holographic' description of the two-dimensional quantum Hall states
in terms of the one-dimensional Calogero particle picture.
Properties of the system can be translated back-and-forth between the two descriptions.
Further connections at the quantum level will be described in
subsequent sections.

\section{The quantum matrix Chern-Simons model}

The properties of the model analyzed in the previous section are classical.
The `states' and `oscillators' that we encountered were due to the \nc nature
of the coordinates and were referring to the classical matrix model.

The full physical content of the model, and its complete equivalence to quantum
Hall (Laughlin) states, is revealed only upon quantization. In fact, some of the
most interesting features of the states, such as filling fraction and quasihole
charge quantization, manifest only in the quantum domain. This will be the subject
of the present section.

\subsection{Quantization of the filling fraction}

The quantization of the Chern-Simons matrix model has been treated in \cite{PolMM}. 
We shall repeat here the basic arguments 
establishing their relevance to the quantum Hall system.

We shall use double brackets for quantum commutators
and double kets for quantum states, to distinguish them
from matrix commutators and $N$-vectors.

Quantum mechanically the matrix elements of $X^i$ become
operators. Since the lagrangian is first-order in time derivatives,
$X^1_{mn}$ and $X^2_{kl}$ are canonically conjugate:
\be
[[ X^1_{mn} , X^2_{kl} ]] = \frac{i}{B} \delta_{ml} \delta_{kn}
\label{Xcanon}
\ee
or, in terms of $A= X^1 + i X^2$
\be
[[ A_{mn} , A_{kl}^\dagger ]] = \frac{1}{B} \delta_{mk} \delta_{nl}
\label{Acanon}
\ee
The hamiltonian, ordered as $\half B\omega \Tr A^\dagger A$, is
\be
H = \sum_{mn} \half B \omega A_{mn}^\dagger A_{mn}
\ee
This is just $N^2$ harmonic oscillators. 
Further, the components of the vector $\Psi_n$
correspond to $N$ harmonic oscillators. Quantized as bosons, 
their canonical commutator is
\be
[[ \Psi_m , \Psi_n^\dagger ]] = \delta_{mn}
\ee

So the system is a priori just $N(N+1)$ uncoupled oscillators.
What couples the oscillators and reduces the system to effectively 
$2N$ phase space variables
(the planar coordinates of the electrons) is the Gauss law constraint
(\ref{G}). In writing it, we in principle encounter operator ordering
ambiguities. These are easily fixed, however, by noting that the
operator $G$ is the quantum generator of unitary rotations of
both $X^i$ and $\Psi$. Therefore, it must satisfy the commutation
relations of the $U(N)$ algebra. The $X$-part is an
orbital realization of $SU(N)$ on the manifold of $N \times N$
hermitian matrices. Specifically, expand $X^{1,2}$ and 
$A , A^\dagger$ in the complete basis of matrices 
$\{1, T^a\}$ where $T^a$ are the $N^2 -1$ normalized
fundamental $SU(N)$ generators:
\be
X^1 = x_0 + \sum_{a=1}^{N^2 -1} x_a T^a ~,~~~ 
\sqrt{B} A =  a_o + \sum_{a=1}^{N^2 -1} a_a T^a 
\ee
$x_a$, $a_a$ are scalar operators. Then, by 
(\ref{Xcanon},\ref{Acanon}) the
corresponding components of $B X^2$ are the conjugate operators
$-i \partial / \partial x_a$, while $a_a , a_a^\dagger$ are
harmonic oscillator operators. We can write the components of
the matrix commutator $G_X = -iB [ X^1 , X^2 ]$ in $G$ 
in the following ordering
\beqar
G_X^a &=& -i f^{abc} x_b \frac{\partial}{\partial x_a} \\
 &=& -i ( A_{mk}^\dagger A_{nk} - A_{nk}^\dagger A_{mk} ) \\
 &=& -i a_b^\dagger f^{abc} a_c 
\eeqar
where $f^{abc}$ are the structure constants of $SU(N)$.
Similarly, expressing $G_\Psi = \Psi \Psi^\dagger$ in the $SU(N)$
basis of matrices, we write its components in the ordering
\be
G_\Psi^a = \Psi_m^\dagger T_{mn}^a \Psi_n
\ee

The operators above, with the specific normal ordering, indeed 
satisfy the $SU(N)$ algebra. The expression of $G_X^a$ in terms
of $x_a$ is like an angular momentum. The expression of 
$G_\Psi^a$ in terms of the oscillators $\Psi_i$ and of
$G_X^a$ in terms of the oscillators $a_a$ is the well-known 
Jordan-Wigner realization of the $SU(N)$
algebra in the Fock space of bosonic oscillators. Specifically,
let $R_{\alpha \beta}^a$ be the matrix elements of the generators
of $SU(N)$ in any representation of dimension $d_R$,
and $a_\alpha , a_\alpha^\dagger$
a set of $d_R$ mutually commuting oscillators. Then the operators
\be
G^a = a_\alpha^\dagger R_{\alpha \beta}^a a_\beta
\ee
satisfy the $SU(N)$ algebra. The Fock space of the oscillators
contains all the symmetric tensor products of $R$-representations
of $SU(N)$; the total number operator of the oscillators identifies
the number of $R$ components in the specific symmetric product.
The expressions for $G_\Psi^a$ and $G_X^a$ are specific cases of 
the above construction for $R^a$ the fundamental ($T^a$) or the
adjoin ($-if^a$) representation respectively.

So, the traceless part of the Gauss law (\ref{G}) becomes
\be
( G_X^a + G_\Psi^a ) |phys \ra\ra =0
\label{SUN}
\ee
where $|phys \ra\ra$ denotes the physical quantum states of the model.
The trace part, on the other hand, expresses the fact that the total
$U(1)$ charge of the model must vanish. It reads
\be
(\Psi_n^\dagger \Psi_n - NB\theta ) |phys \ra\ra = 0
\label{U1}
\ee

We are now set to derive the first nontrivial quantum mechanical
implication: the inverse-filling fraction is quantized to integer
values. To see this, first notice that the first term in (\ref{U1})
is nothing but the total number operator for the oscillators $\Psi_n$
and is obviously an integer. So we immediately conclude that $NB\theta$
must be quantized to an integer. 

However, this is not the whole story. Let us look again at the
$SU(N)$ Gauss law (\ref{SUN}). It tells us that physical states
must be in a singlet representation of $G^a$. The orbital part
$G_X^a$, however, realizes only representations arising out of
products of the adjoin, and therefore it contains only irreps
whose total number of boxes in their Young tableau is an integer 
multiple of $N$.  Alternatively, the $U(1)$ and $Z_N$ part of $U$ 
is invisible in the transformation $X^i \to U X^i U^{-1}$ and thus
the $Z_N$ charge of the operator realizing this transformation
on states must vanish. (For instance, for $N=2$,
$G^a$ is the usual orbital angular momentum in 3 dimensions which
cannot be half-integer.)

Since physical states are invariant under the sum of $G_X$ and
$G_\Psi$, the representations of $G_\Psi$ and $G_X$ must be 
conjugate to each other so that their product contain
the singlet. Therefore, the irreps of $G_\Psi$ must also have 
a number of boxes which is a multiple of $N$. 
The oscillator realization (\ref{U1}) contains all the
symmetric irreps of $SU(N)$, whose Young tableau consists of a single
row. The number of boxes equals the total number operator of the
oscillators $\Psi_n^\dagger \Psi_n$. So we conclude that $NB\theta$
must be an integer multiple of $N$ \cite{PolMM}, that is,
\be
B\theta = \frac{1}{\nu} = k ~,~~~ k={\rm integer}
\label{nuq}
\ee

The above effect has a purely group theoretic origin. The same
effect, however, can be recovered using topological considerations,
by demanding invariance of the quantum action $exp(iS)$ under gauge
$U(N)$ transformations with a nontrivial winding in the temporal
direction \cite{PolMM}. This is clearly the finite-$N$ counterpart
of the level  quantization for the noncommutative Chern-Simons term
as exposed in a previous section, namely $4\pi \lambda = {\rm integer}$.
By (\ref{lambda}) this is equivalent to (\ref{nuq}).

By reducing the model to the dynamics of the eigenvalues of
$X^1$ we recover a quantum Calogero model with hamiltonian
\be
H = \sum_{n=1}^N \left( \frac{\omega}{2B} p_n^2 + \frac{B\omega}{2}
x_n^2 \right) 
+ \sum_{n\neq m} \frac{k(k+1)}{(x_n - x_m )^2}
\label{Hqcalo}
\ee 
Note the shift of the coupling constant from $k^2$ to $k(k+1)$
compared to the classical case. This is a quantum reordering effect
which results in the shift of $\nu^{-1}$ from $k$ to $k+1 \equiv n$.
The above model is, in fact, perfectly well-defined even for 
fractional values of $\nu^{-1}$, while the matrix model that
generated it requires quantization. This is due to the fact that,
by embedding the particle system in the matrix model, we have
augmented its particle permutation symmetry $S_N$ to general 
$U(N)$ transformations; while the smaller symmetry $S_N$ is 
always well-defined, the larger $U(N)$ symmetry becomes anomalous 
unless $\nu^{-1}$ is quantized.

\subsection{Quantum states}

We can now examine the quantum excitations of this theory.
The quantum states of the model are simply states in the Fock
space of a collection of oscillators. The total energy is the
energy carried by the $N^2$ oscillators $A_{mn}$ or $a_a$.
We must also impose the constraint (\ref{SUN}) and (\ref{U1})
on the Fock states. Overall, this becomes a combinatorics
group theory problem which is in principle doable, although
quite tedious.

Fortunately, we do not need to go through it here. The quantization
of this model is known and achieves its most intuitive description
in terms of the states of the corresponding Calogero model. We
explain how.

Let us work in the $X^1$ representation, $X^2$ being its canonical
momentum. Writing $X^1 = U \Lambda_1 U^{-1}$ with $\Lambda =
{diag} \,\{x_i \}$ being its eigenvalues, we can view the state of the 
system as a wavefunction of $U$ and $x_n$. The gauge generator
$G_X^a$ appearing in the Gauss law (\ref{SUN}) is actually the
conjugate momentum to the variables $U$. Due to the Gauss law,
the angular degrees of freedom $U$ are constrained to be in a
specific angular momentum state, determined by the representation 
of $SU(N)$ carried by the $\Psi_n$.
{}From the discussion of the previous section,
we understand that this is the completely symmetric representation
with $nN = N/\nu$ boxes in the Young tableau. So the dynamics of
$U$ are completely fixed, and it suffices to consider the states
of the eigenvalues. These are described by the states of the 
quantum Calogero model. The hamiltonian of the Calogero model 
corresponds to the matrix potential $V = \half B\omega \Tr (X^i)^2$,
which contains all the relevant information for the system.

Calogero energy eigenstates are expressed in terms of $N$
positive, integer `quasi-occupation numbers' $n_j$ (quasinumbers,
for short), with the property
\be
n_j - n_{j-1} \geq n = \frac{1}{\nu} ~,~~~j=1,\dots N
\label{repul}
\ee
In terms of the $n_j$ the spectrum becomes identical to the
spectrum of $N$ independent harmonic oscillators
\be
E = \sum_{j=1}^N E_j = \sum_{j=1}^N \omega \left( n_j +\half\right)
\ee
The constraint (\ref{repul}) means that the $n_j$ cannot be packed 
closer than $n=\nu^{-1}$, so they have a `statistical repulsion' 
of order $n$.
For filling fraction $\nu=1$ these are ordinary fermions, while
for $\nu^{-1} = n>1$ they behave as particles with an enhanced
exclusion principle. 

The scattering phase shift between Calogero
particles is $exp(i\pi/\nu)$. So, in terms of the phase that
their wavefunction picks upon exchanging them, they look like
fermions for odd $n$ and bosons for even $n$ \cite{PolFS}. 
Since the underlying
particles (electrons) must be fermions, we should pick $n$ odd.

The energy `eigenvalues' $E_j$ are the quantum 
analogs of the eigenvalues of the matrix $\half B\omega (X^i)^2$.
The radial positions $R_j$ are determined by
\be
\half B\omega R_j^2 = E_j ~~~\rightarrow ~~~
R_j^2 = \frac{2 n_j +1 }{B}
\ee
So the quasinumbers $2n_j +1$ determine the radial positions 
of electrons. The ground state values are the smallest 
non-negative integers satisfying (\ref{repul})
\be
n_{j,gs} = n(j-1) ~,~~~ j=1 , \dots N
\label{gs}
\ee
They form a `Fermi sea' but with a density of states dilated
by a factor $\nu$ compared to standard fermions. This state
reproduces the circular quantum Hall droplet. Its radius maps
to the Fermi level, $R \sim \sqrt{(2n_{N,gs} +1)/B} \sim
\sqrt{2N\theta}$.

Quasiparticle and quasihole states are identified in a way 
analogous to particles and holes of a Fermi sea.
A quasiparticle state is obtained by peeling a `particle' from
the surface of the sea (quasinumber $n_{N,gs}$) and putting it to a
higher value ${n'}_N > n(N-1)$. This corresponds to an electron
in a rotationally invariant state at radial position $R' \sim
\sqrt{2({n'}_N +1)/B}$. Successive particles can be excited this
way. The particle number is obviously quantized
to an integer (the number of excited quasinumbers) and we can
only place them outside the quantum Hall droplet.

Quasiholes are somewhat subtler: they correspond to the minimal
excitations of the ground state {\it inside} the quantum Hall
droplet. This can be achieved by leaving all quasinumber $n_j$
for $j \leq k$ unchanged, and increasing all $n_j$, $j>k$ by one
\beqar
n_j &=& n(j-1) ~~~~ j \leq k \\
    &=& n(j-1)+1 ~~~ k < j \leq N 
\eeqar
This increases the gap between $n_k$ and $n_{k+1}$ to $n+1$
and creates a minimal `hole.' 

This hole has a particle number $-q = -1/n = -\nu$. 
To see it, consider removing a particle altogether from 
quasinumber $n_k$. This would create a gap of $2n$ between 
$n_{k-1}$ and $n_{k+1}$. The extra gap $n$ can be considered 
as arising out of the formation of $n$ holes (increasing $n_j$ 
for $j\geq k$ $n$ times). Thus the absence of a particle 
corresponds to $n$ holes. We therefore
obtain the important result that the quasihole charge is 
naturally quantized to units of
\be
q_h = \nu = \frac{1}{n}
\label{holeq}
\ee
in accordance with Laughlin theory.

We conclude by stressing once more that there is no fundamental
distinction between particles and holes for finite $N$. A particle
can be considered as a nonperturbative excitation of many holes
near the Fermi level, while a hole can be viewed as a coherent
state of many particles of minimal excitation.

\subsection{Final remarks on the matrix model}

The quantization of the inverse filling fraction
and, importantly, the quasihole charge quantization emerged as 
quantum mechanical consequences of this model. 
The quantizations of the two parameters had a rather different
origin. We can summarize here the basic meaning of each:

Quantization of the inverse filling fraction is basically
angular momentum quantization. The matrix commutator of 
$[X_1 , X_2 ]$ is an orbital angular momentum in the compact
space of the angular parameters of the matrices, and it must
be quantized. Alternatively (and equivalently), it can be
understood as a topological quantization condition due to a
global gauge anomaly of the model.

Quantization of the quasihole charge, on the other hand,
is nothing but harmonic
oscillator quantization. Quasiholes are simply individual quanta
of the oscillators $A_{mn}$. The square of the radial coordinate
$R^2 = (X^1)^2 + (X^2)^2$ is basically a harmonic oscillator. 
$\sqrt{B} X_1$ and $\sqrt{B} X_2$ are canonically conjugate,
so the quanta of $R^2$ are $2/B$. Each quantum increases $R^2$
by $2/B$ and so it increases the area by $2\pi/B$. This creates
a charge deficit $q$ equal to the area times the ground state 
density $q=( 2\pi/B ) \cdot (1/2\pi\theta) = 1/\theta B = \nu$.
So the fundamental quasihole charge is $\nu$.

An important effect, which can be both interesting and frustrating,
is the quantum shift in the effective value of the inverse filling
fraction from $k$ to $n=k+1$. This is the root of the famous
fermionization of the eigenvalues of the matrix model in the
singlet sector ($k=0$). Its presence complicates some efforts
to reproduce layered quantum Hall states, as it frustrates the
obvious charge density counting.

There are many questions on the above model that we left untouched, some of them
already addressed and some still open \cite{HR}-\cite{GMSS}. Their list includes
the description of Hall states with spin, the treatment of cylindrical,
spherical or toroidal space topologies, the description of states with
nontrivial filling fraction, the exact mapping between quantities of physical
interest in the two descriptions, the inclusion of electron interactions etc.
The interested reader is directed to the numerous papers in the literature
dealing with these issues. In the concluding section we prefer to present
an alternative \nc fluid description for quantum many-body states.

\section{The \nc Euler picture and Bosonization}

In the previous sections we reviewed the \nc picture of the Lagrange formulation of
fluids and its use in the quantum Hall effect. The Euler formulation, on the other hand,
was peculiar in that it allowed for a fully commutative description,
leading to the Seiberg-Witten map.

This, however, is not the only possibility. Indeed, we saw that there were {\it two}
potential descriptions for the density of the fluid, one inherently commutative
(\ref{rhok}) and one inherently \nc (\ref{ncrhoXX}). Although the commutative one
was adopted, one could just as well work with the \nc one, expecting to recover
the standard Euler description only at the commutative limit. As it turns out, this is
a very natural description of fluids consisting of fermions. Since the \nc density is an
inherently bosonic field, it affords a description of fermionic systems in terms of
bosonic field variables, naturally leading to bosonization.

\subsection{Density description of fermionic many-body systems}

The starting point will be a system of $N$ non-interacting fermions in $D=1$ spatial
dimensions. The restriction of the dimensionality of space at this point is completely
unnecessary and inconsequential, and is imposed only for conceptual and notational simplification
and easier comparison with previous sections. In fact, much of the formalism will not even
make specific reference to the dimensionality of space.

We shall choose our fermions to be noninteracting and carrying no internal degrees of
freedom such as spin, color etc. (there is no conflict with the spin-statistics theorem in this
first-quantized, many-body description). Again, this is solely for convenience and to allow us to
focus on the main conceptual issue of their fluid description rather than other dynamical
questions. The only remaining physical quantity is the single-particle hamiltonian defining
their dynamics, denoted $H\sp (x,p)$. 
Here $x,p$ are single-particle coordinate and momentum operators, together forming a `\nc plane', 
with the role of $\theta$ played by $\hbar$ itself:
\be
[ x , p ]\sp = i \hbar
\ee
The subscript sp will be appended to single-particle operators or relations (except $x$ and $p$) 
to distinguish them from upcoming field theory quantities.

Single-particle states are elements of the irreducible representation of the above Heisenberg
commutator. A basis would be the eigenstates $|n\ra$ of $H\sp$ corresponding to eigenvalues $E_n$
(assumed nondegenerate for simplicity). The states of the $N$-body system, on the other hand,
are fully antisymmetrized elements of the $N$-body Hilbert space
consisting of $N$ copies of the above space. They can be expressed in a Fock description in terms
of the occupation number basis $N_n = 0,1$ for each single particle level.
The ground state, in particular, is the state $|1, \dots 1, 0 , \dots \ra$ with the $N$ lowest
levels occupied by fermions.

An alternative description, however, working with a {\it single} copy of the above space is possible,
in terms of a single-particle density-like operator \cite{Sakita,KaNa}. Specifically, define the
(hermitian) single-particle operator $\rho$ whose eigenvalues correspond to the
occupation numbers $N_i = 1$ for a set of $N$ specific filled single-particle states and $N_i = 0$
for all other states:
\be
\rho = \sum_{n=1}^N |\psi_n \ra \la \psi_n |
\ee
Clearly $\rho$ is a good description of the $N$-body fermion system whenever the fermions occupy
$N$ single-particle states. The ground state $\rho_0$, in particular, is such a state and would
correspond to
\be
\rho_0 = \sum_{n=1}^N |n\ra \la n|
\ee
Due to the Schr\"odinger evolution of the single-particle states $|n\ra$,
the operator $\rho$ satisfies the evolution equation
\be
i\hbar {\dot \rho} = [H\sp , \rho ]\sp
\label{drho}
\ee
Here and for the rest of this chapter we shall display $\hbar$ explicitly, as a useful tool to
keep track of scales.

This description has several drawbacks. It is obviously limited from the fact that it can describe
only `factorizable' states, that is, basis states in some appropriate Fock space, but not their
linear combinations (`entangled' states).
This is serious, as it violates the quantum mechanical superposition principle,
and makes it clear that this cannot be a full quantum description of the system. Further, the
operator $\rho$ must be a projection operator with exactly $N$ eigenvalues equal to one and the
rest of them vanishing, which means that it must satisfy the algebraic constraint
\be
\rho^2 = \rho ~,~~~ \Tr \rho = N
\ee
So $\rho$ is similar to the density matrix, except for its trace.

In spite of the above, we shall see that this is a valid starting point for a full description of the
many-body quantum system in a second-quantized picture. To give $\rho$ proper dynamics, we must write 
an action that leads to the
above equations (evolution plus constraints) in a canonical setting. The simplest way to achieve this
is by `solving' the constraint in terms of a unitary field $U$ as:
\be
\rho = U^{-1} \rho_0 U
\label{rhoU}
\ee
with $\rho_0$ the ground state. Any $\rho$ can be expressed as above, $U$ being a unitary operator
mapping the first $N$ energy eigenstates to the actual single-particle states entering the definition
of $\rho$. An appropriate action for $U$ is
\be
S = \int dt \Tr \left( i\hbar \rho_0 {\dot U} U^{-1} - U^{-1} \rho_0 U H\sp \right)
\label{Srho}
\ee
It is easy to check that it leads to (\ref{drho}) for (\ref{rhoU}).
Note that the first term in the action is a first-order kinetic term, defining a canonical one-form. 
The matrix elements of $U$, therefore, encode both coordinates and momenta and constitute
the full phase space variable of the system. The Poisson brackets of $U$ and, consequently, $\rho$
can be derived by inverting the above canonical one-form. The result is that the matrix elements
$\rho_{mn}$ of $\rho$ have Poisson brackets
\be
\{ \rho_{m_1 n_1} , \rho_{m_2 n_2} \} = \frac{1}{i\hbar} (\rho_{m_1 n_2} \delta_{m_2 n_1}
- \rho_{m_2 n_1} \delta_{m_1 n_2} )
\label{rrclass}
\ee
The second term in the action is the Hamiltonian $H = \Tr (\rho H\sp )$ and represents the sum of the
energy expectation values of the $N$ fermions.

\subsection{The correspondence to a \nc fluid}

It should be clear the the above description essentially defines a \nc fluid. Indeed, the
operators $U$ and $\rho$ act on the Heisenberg Hilbert space and can be expressed in terms of
the fundamental operators $x,p$. As such, they are \nc fields. The constraint for $\rho$ is
the \nc version of the relation $f^2 = f$ defining the characteristic function of a domain.
We can, therefore, visualize $\rho$ as a `droplet' of a \nc fluid that fills a `domain' 
of the \nc plain with a droplet 
`height' equal to 1. The actual density of the fluid is fixed by the integration formula on the
\nc plane, assigning an area of $2\pi \hbar$ to each state on the Hilbert space. So the value
of the density inside the droplet becomes $1/2\pi \hbar$.

A similar picture is obtained by considering the classical `symbol' of the above operator, using
the Weyl-ordering mapping. The corresponding commutative function represents a droplet with a fuzzy
boundary (the field drops smoothly from 1 to 0, and can even become negative at some points), 
but the bulk of the droplet and its exterior are at constant density (0 or 1).

As one should expect, this is the value of the density of states on phase space according to the
semiclassical quantization condition assigning one quantum state per phase space area $h = 2\pi \hbar$.
The above description is the quantum, fuzzy, \nc analog of the classical phase space density. According
to the Liouville theorem, a collection of particles with some density on the phase space evolves
in an area-preserving way, so a droplet of constant density evolves into a droplet of different shape
but the same constant density \cite{APdr}.

The ground state $\rho_0$ corresponds to a droplet filling a `lake' in phase space in which the classical
value of the single particle energy satisfies
\be
H\sp (x,p) \le E_F
\ee
This ensures the minimal energy for the full state. The boundary of the droplet is at the line
defined by the points $H\sp = E_F$, the highest energy of any single particle. This is the Fermi energy. 

The unitary transformation $U$ maps to a `star-unitary' commutative function satisfying $U * U^* = 1$.
One could think that in the commutative (classical) limit it becomes a phase, $U = {\rm exp}[i \phi (x,p)]$.
This, however, is not necessarily so. $U$ enters into the definition of $\rho$ only through the adjoin
action $\rho = U^* * \rho_0 * U$. If $U$ became a phase in the commutative limit, it would give
$\rho = \rho_0$ (upon mapping star products to ordinary products), creating no variation. 
The trick is that $U(x,p)$ can contain terms of order $\hbar^{-1}$: since the star-products in the
definition of $\rho$ in terms of $U$ reproduce $\rho_0$ plus terms of order $\hbar$, the overall
result will be of order $\hbar^0$ and remain finite in the classical limit. So $U(x,p)$ may not map
to a finite function in this limit; its action on $\rho_0$, however, is finite and defines a canonical 
transformation, changing the shape of the droplet. Overall, we have a correspondence
with a fuzzy, incompressible phase space fluid in the density (Euler) description.

\subsection{Quantization and the full many-body correspondence}

What makes this description viable and useful is that it reproduces the {\it full Hilbert space} 
of the $N$ fermions upon quantization.

The easiest way to see this is to notice that the action (\ref{Srho}) is of the Kirillov-Kostant-Souriau
form for the group of unitary transformations on the Hilbert space. For concreteness, we may introduce
a cutoff and truncate the Hilbert space to the $K$ first energy levels $K \gg N$. Then the above becomes
the KKS action for the group $U(K)$. Its properties and quantization are fully known, and we summarize
the basic points.

Both $\rho = U^{-1} \rho_0 U$ and the action (\ref{Srho}) are invariant under time-dependent transformations
\be
U(t) \to V(t) U(t) ~,~~~ [\rho_0 , V(t) ] =0
\ee
for any unitary operator ($K \times K$ unitary matrix) commuting with $\rho_0$. This means that the
corresponding `diagonal' degrees of freedom of $U$ are redundant and correspond to a gauge invariance
of the description in terms of $U$. This introduces a Gauss law as well as a `global gauge anomaly'
for the action that requires a quantization condition, akin to the magnetic monopole quantization or
level quantization for the Chern-Simons term. The end result is:

$\bullet$ The eigenvalues of the constant matrix $\rho_0$ must be integers for a consistent quantization.

On the other hand, the classical Poisson brackets for $\rho$ (\ref{rrclass}) become, upon quantization,
\be
[[ \rho_{m_1 n_1} , \rho_{m_2 n_2} ]] = \rho_{m_1 n_2} \delta_{m_2 n_1}
- \rho_{m_2 n_1} \delta_{m_1 n_2}
\label{rrquant}
\ee
where we used, again, double brackets for quantum commutators to distinguish from matrix (single-particle)
commutators. The above is nothing but the $U(K)$ algebra in a `cartesian' basis (notice how $\hbar$ has
disappeared). The quantum Hilbert space, therefore, will form representations of $U(K)$. The Gauss law,
however, imposes constraints on what these can be. The end result is:

$\bullet$ The quantum states form an irreducible representation of $U(K)$ determined by a Young tableau
with the number of boxes in each row corresponding to the eigenvalues of $\rho_0$.

In our case, the eigenvalues are $N$ 1s and $K-N$ 0s, already properly quantized. So the Young tableau 
corresponds to a single
column of $N$ boxes; that is, the $N$-fold fully antisymmetric representation of $U(K)$.

This is exactly the Hilbert space of $N$ fermions on $K$ single-particle states! The dimensionality
of this representation is
\be
D = \frac{K!}{N! (K-N)!}
\ee
matching the number quantum states of $N$ fermions in $K$ levels. The matrix elements of the operator
$\rho_{mn}$ in the above representation can be realized in a Jordan-Wigner construction involving $K$
fermionic oscillators $\Psi_n$, as
\be
\rho_{mn} = \Psi_n^\dagger \Psi_m
\ee
satisfying the constraint
\be
\sum_{n=1}^K \Psi_n^\dagger \Psi_n = N
\ee
This $\Psi$ is essentially the second-quantized fermion field, the above relation being the constraint
to the $N$-particle sector. The quantized hamiltonian operator for $\rho$ in this realization becomes 
\be
H = \Tr (\rho H\sp) = \sum_{m,n} \Psi_m^\dagger (H\sp )_{mn} \Psi_n
\ee
and thus also corresponds to the second-quantized many-body hamiltonian. 
Overall, this becomes a {\it complete}
description of the many-body fermion system in terms of a quantized \nc density field $\rho$
or, equivalently, the unitary \nc field $U$.

It is worth pointing out that in the limit $K \to \infty$ the algebra (\ref{rrquant}) becomes
infinite-dimensional and reproduces the so-called ``$W_\infty$ algebra. This algebra has a host
of representations, one of which corresponds to the Hilbert space of $N$ fermions. In the finite
$K$ case the conditions $\rho^2 = \rho$ and $\tr \rho = N$ fixed the Hilbert space. Similar
conditions, corresponding to the appropriate choice of a `vacuum' (highest-weight) state, fix
the desired representation of the $W_\infty$ density algebra.

The commutative limit of the above
algebra, on the other hand, corresponds to the standard Poisson brackets of phase space density
functions, as implied by the underlying canonical structure of $x$ and $p$. We observe that the 
quantization of the algebra involves two steps: the standard step of turning Poisson brackets into
($1/i\hbar$ times) quantum commutators, as well as a {\it deformation} of the Poisson structure
(see \cite{GJPP} for an elaboration). This will be crucial in the upcoming discussion.

\subsection{Higher-dimensional \nc bosonization}

The above also constitutes an exact bosonization of the fermion system. Indeed, the fields $\rho$
or $U$ are bosonic, so they afford a description of fermions without use of Grassman variables.
The price to pay is the increase of dimensionality (two phase space rather than one space dimensions)
and the \nc nature of the classical $\rho$-dynamics, even before quantization.

The correspondence to traditional bosonization can be achieved through the Seiberg-Witten map
on the field $U$.
We shall not enter into any detail here, but the upshot of the story is that the action
(\ref{Srho}) maps to the (commutative) action of a one-dimensional chiral boson under this
map. The corresponding space derivative of the field is an abelian `current' that maps to the
boundary of the classical fluid droplet, which parametrizes the full shape of the fluid.
Overall this recovers standard abelian bosonization results \cite{Boso} in the \nc hydrodynamic setting.
Generalizations to particles carrying internal degrees of freedom are possible and lead to
the Wess-Zumino-Witten action for nonabelian bosonization \cite{WZW}.

Most intriguingly, much of the above discussion can be exported to higher dimensions.
The formalism extends naturally to higher dimensions, the matrix $\rho$ now acting on the
space of states of a single particle in $D$ spatial dimensions. The crucial difference,
however, is that the Seiberg-Witten map of the higher-dimensional action yields a nontrivial
action in $2D$ (phase space) dimensions that, unlike the $D=1$ case, does not reduce to a $D$-dimensional
chiral boson action.

We can obtain a more economical description
by performing the Seiberg-Witten map only on a two-dimensional \nc subspace, leaving the rest of the 
$2D$-dimensional space untouched. This transformation works similarly to the $D=1$ case, leading to a
description in terms of a field in one residual (commutative) dimension as well as the remaining $2D-2$
noncommutative ones. This constitutes a `minimal' bosonization in the \nc field theory setting \cite{APbos}.
(For other approaches on higher dimensional bosonization see \cite{HiBoz}.)

The form of the above theory can be motivated by starting with the fully classical, commutative
picture of our density droplet in phase space of constant density $\rho_0 = 1/(2\pi \hbar)^D$,
whose shape is fully determined in terms of its boundary.
A convenient way to parametrize the boundary is in terms of the value of one of the phase space
coordinates, say $p_{_D}$, on the boundary as a function of the $2D-1$ remaining ones. We write
\be
\left. p_{_D} \right|_{\rm boundary} \equiv R(x_1 , p_1 ; \dots x_{_D} )
\ee
$R$ will be the boundary field of the theory. For notational convenience, we rename the variable
conjugate to the eliminated variable $p_{_D}$ (that is, $x_{_D}$) $\sigma$ and write $\phi^\alpha$
($\alpha = 1, \dots 2D-2$) for the
remaining $2D-2$ phase space dimensions $(x_n , p_n )$ ($n=1, \dots D-1$).

The dynamics of the classical system are determined by the canonical Poisson
brackets of the field $R(\sigma,\phi )$. These can be derived through a hamiltonian reduction
of the full density Poisson brackets on the phase space \cite{APdr} and we simply quote the result.
We use $\thab = \{ \phi^\alpha , \phi^\beta \}_{\rm sp}$ for the standard (Darboux) single-particle Poisson 
brackets of $\phi$ (that is, $\thab = \epsilon^{\alpha \beta}$ if $\alpha$ and $\beta$ correspond
to $x_n$ and $p_n$, otherwise zero),
as well as the shorthand $R_{1,2} = R(\sigma_{1,2} , \phi_{1,2} )$, with $1$ and $2$ labeling the
two points in the $2D-1$ dimensional space $(\sigma , \phi )$ at which we shall calculate the
brackets. The field theory Poisson brackets for $R_1$ and $R_2$ read, in an obvious notation:
\be
\{ R_1 , R_2 \} = \frac{1}{\rho_0}
\left[ - \delta' (\sigma_1 -\sigma_2) \, \delta (\phi_1 - \phi_2)
- \delta (\sigma_1 -\sigma_2 )\,
\{ R_1 , \delta (\phi_1 - \phi_2 ) \}_{\rm sp1} \right]
\label{PR}
\ee
Similarly, the hamiltonian for the field $R$ is
the integral of the single-particle hamiltonian over the bulk of the droplet and reads
\be
H = \rho_o \int dp_{_D} \, d\sigma \, d^{2d}\phi H\sp (\sigma , \phi) \, \vartheta (R-p_{_D} )
\label{H}
\ee
where $\vartheta (x) = \half [1+ \rm{sgn} (x)]$ is the step function.
(\ref{PR}) and (\ref{H}) define a bosonic field theory
(in a hamiltonian setting) that describes the droplet classically.

The correct quantum version of the theory {\it cannot} simply be obtained by
turning the above Poisson brackets into quantum commutators. We have already
encountered a similar situation in the previous subsection: the commutative, classical
Poisson algebra of the density operator $\rho$ is deformed into the $W_\infty$ algebra
(or its finite $U(K)$ truncation) in the quantum case.

This observation will guide us in motivating the correct quantum commutators for the boundary field.
We observe that the first, $R$-independent term of the above Poisson brackets
reproduces a current algebra in the $\sigma$-direction, exactly as in one-dimensional
bosonization. The second, homogeneous term, on the other hand, has the form of a
density algebra in the residual $2D-2$ phase space dimensions. In this sense, the
field $R$ is partly current and partly density. Taking our clues from
standard bosonization and the story of the previous subsections, we propose that the
current algebra part remains undeformed upon quantization, while the density part
gets deformed to the corresponding \nc structure. A simple way to do that and still use
the same (commutative) phase space notation is in the $*$-product language.
Specifically, we turn the single-particle Poisson brackets to noncommutative Moyal brackets
$\{ . , . \}_*$ on the $2d$-dimensional phase space manifold $\phi^\alpha$. The full
deformed field theory Poisson brackets, now, read:
\be
\{ R_1 , R_2 \} = \frac{1}{\rho_0}
\left[ -\delta' (\sigma_1 -\sigma_2 ) \delta (\phi_1 - \phi_2 )
- \delta (\sigma_1 -\sigma_2 )
\{ R_1 , \delta (\phi_1 - \phi_2 ) \}_{*1} \right]
\label{qPB}
\ee
The Moyal brackets between two functions of $\phi$ are expressed
in terms of the noncommutative Groenewald-Moyal star-product on the phase
space $\phi$ \cite{star}:
\be
\{ F(\phi) , G(\phi) \}_* = \frac{1}{i\hbar} \left[
F(\phi) * G(\phi) - G(\phi) * F(\phi) \right]
\label{Moy}
\ee
with $\hbar$ itself being the noncommutativity parameter.
Correspondingly, the hamiltonian $H$ is given by expression (\ref{H})
but with $*$-products replacing the usual products between its terms.

The transition to the matrix (`operator') notation can be done in the standard
way, as exposed in the introductory sections, by choosing any basis of states
$\psi_a$ in the single-particle Hilbert space. This would map the field $R(\sigma,\phi)$ to
dynamical matrix elements $R^{ab} (\sigma)$. 
The only extra piece that we need is the matrix representation of the delta
function $\delta (\phi_1 - \phi_2 )$, with defining property
\be
\int d^{2d} \phi_1 F(\phi_1 ) \, \delta (\phi_1 - \phi_2 ) = F(\phi_2 )
\ee
Since $\delta (\phi_1 - \phi_2 )$ is a function of two variables, its
matrix transform in each of them will produce a symbol with four indices
$\delta^{a_1 b_1 ; a_2 b_2}$. The above defining relation in
the matrix representation becomes
\be
(2\pi\hbar)^{(D-1)} F^{a_1 b_1} \delta^{b_1 a_1 ; a_2 b_2} = F^{a_2 b_2}
\ee
which implies
\be
\delta^{a_1 b_1 ; a_2 b_2} = \frac{1}{(2\pi\hbar)^{(D-1)}} \delta^{a_1 b_2}
\delta^{a_2 b_1}
\ee
With the above, and using $\rho_o = 1/(2\pi\hbar)^D$,
the canonical Poisson brackets of the matrix $R^{ab}$
become
\be
\{ R_1^{ab} , R_2^{cd} \} = - 2\pi\hbar  \delta'
(\sigma_1 -\sigma_2 ) \delta^{ad} \delta^{cb}
+ 2\pi i \delta (\sigma_1 -\sigma_2 )
(R_1^{ad} \delta^{cb} - R_1^{cb} \delta^{ad} ) 
\label{alga}
\ee
Not surprisingly, we recover a structure for the second term similar to the
one for $\rho$ of the previous subsection as expressed in (\ref{rrquant}).

We are now ready to perform the quantization of the theory. The fields $R^{ab} (\sigma)$ 
become operators whose quantum commutator is given by the above Poisson brackets times
$i\hbar$. Defining, further, the Fourier modes
\be
R_k^{ab} = \int \frac{d\sigma}{2\pi\hbar} R^{ab}(\sigma) e^{-ik\sigma}
\label{Fou}
\ee
the quantum commutators become 
\be
[[ R_k^{ab} , R_{k'}^{cd} ]] = k \delta (k+k' ) \delta^{ad} \delta^{cb}
- R_{k+k'}^{ad} \delta^{cb} + R_{k+k'}^{cb} \delta^{ad}
\label{quantcom}
\ee
The zero mode $R_0^{aa} \equiv N$ is a Casimir and represents the total fermion number.
For a compact dimension $\sigma$, normalized to a circle of length $2\pi$, the Fourier
modes become discrete.

The above is also recognized as a chiral current algebra for the matrix field $R_k^{ab}$
on the unitary group of transformations of the first-quantized states $\psi_a$.
To make this explicit, consider again the finite-dimensional truncation of the
Hilbert space into $K$ single-particle states; that is, $a,b,c,d=1,\dots K$ (this
would automatically be the case for a compact phase space $\{ \phi^\alpha \}$).
As remarked before, the homogeneous part of the above commutator is the $U(K)$
algebra in a `cartesian' parametrization. To bring it into the habitual form,
define the hermitian $K \times K$ fundamental generators of
$U(K)$, $T^A$, $A=0,\dots K^2 -1$,
normalized as $\tr (T^A T^B ) = \half \delta^{AB}$,
which fix the $U(M)$ structure constants $[ T^A , T^B ] = i f^{ABC} T^C$
(with $f^{0AB} = 0$). Using the $T^A$ as a basis we express the quantum
commutators (\ref{quantcom}) in terms of the $R^A = \tr (T^A R)$ as
\be
[[ R_k^A , R_{k'}^B ]] = \half k \delta (k+k' ) \delta^{AB} +
i f^{ABC} R_{k+k'}^C
\ee
This is the so-called Kac-Moody algebra for the group $U(K)$.

The coefficient $k_{_{KM}}$ of the central extension of the Kac-Moody algebra
(the first, affine term) must be quantized to an integer to have
unitary representations. Interestingly, this coefficient in the above
commutators emerges quantized to the value $k_{_{KM}}=1$.
This is crucial for bosonization \cite{WZW}. The $k_{_{KM}}=1$ algebra has a 
unique irreducible
unitary representation over each `vacuum'; that is, over highest
weight states annihilated by all $R^A (k)$ for $k>0$ and transforming
under a fully antisymmetric $SU(K)$ representations under $T^A (0)$.
These Fock-like representations correspond exactly
to the perturbative Hilbert space of excitations of the many-body
fermionic system over the full set of possible Fermi sea ground states.
The $U(1)$ charge $R_0^0$, which is a Casimir, corresponds to the total
fermion number; diagonal operators $R_k^H $, for $k<0$ and $H$ in the
Cartan subgroup of $U(K)$ generate `radial' excitations
in the Fermi sea along each direction in the residual phase space variables;
while off-diagonal operators $R_k^T $, for $k<0$ and
$T$ off the Cartan subgroup, generate transitions of fermions between
different points of the Fermi sea.

In the above, we have suddenly introduced the word `perturbative' in the
mapping between states of the field $R$ and many-body fermion states. We had
started with a full, nonperturbative description of the system before we
reduced it to boundary variables. Where did perturbative come from?

This is a standard feature of bosonization, true also in the one-dimensional
case. The boundary of the droplet could in principle `hit' upon itself,
breaking the droplet into disconnected components. The field $R$ in such cases would
develop `shock waves' and lose single-valuedness. Quantum mechanically,
the above situation corresponds to locally depleting the Fermi sea. This is an essentially
nonperturbative phenomenon, whose account would require the introduction of
branches for the field $R$ after the formation of shock waves and corresponding
boundary conditions between the branches. Quantum mechanically it would require
nontrivial truncations and identifications of states in the Hilbert space of the
quantum field $R$. In the absence of that, the bosonic theory gives an exact description
of the Fermi system up to the point that the Dirac sea would be depleted.
This is adequate for many-body applications.
 
Finally, the hamiltonian of the bosonic theory becomes
\be
H = \int \frac{dp_{_D} d\sigma}{2\pi\hbar} \tr H\sp (\sigma , p_{_D}, {\hat \phi})
\vartheta (R-p_{_D} )
\label{HQ}
\ee
where $p_{_D}$ remains a scalar integration parameter while $\hat \phi$
become (classical) matrices and $R$ is an operator matrix field as before.
Clearly there are issues of ordering in the above expression, 
matrix (\nc) as well as quantum, just as in standard $1+1$-dimensional bosonization.

To demonstrate the applicability of this theory we shall work out
explicitly the simplest nontrivial example of higher-dimensional
bosonization: a system of $N$ noninteracting two-dimensional fermions
in a harmonic oscillator potential. The single-particle
hamiltonian is
\be
H\sp = \half (p_1^2 + x_1^2 + p_2^2 + x_2^2 )
\ee
For simplicity we chose the oscillator to be isotropic
and of unit frequency. The single-body spectrum is the direct sum of two
simple harmonic oscillator spectra, $E_{mn} = \hbar (m+n+1)$, $m,n=0,1,\dots$.
Calling $m+n+1=K$, the energy levels are $E_K = \hbar K$ with degeneracy $K$.

The $N$-body ground state consists of fermions filling states $E_K$
up to a Fermi level $E_F = \hbar K_F$. In general, this state is degenerate,
since the last energy level of degeneracy $K_F$ is not fully occupied.
Specifically, for a number of fermions $N$ satisfying
\be
N = \frac{K_F (K_F -1)}{2} + M ~,~~~ 0 \le M \le K_F
\ee
the Fermi sea consists of a fully filled bulk (the first term above)
and $M$ fermions on the $K_F$-degenerate level at the surface.
The degeneracy of this many-body state is 
\be
g (K_F , M) = \frac{K_F !}{M! (K_F -M)!}
\ee
representing the ways to distribute the $M$ last fermions over $K_F$ states,
and its energy is
\be
E (K_F , M) = \hbar \frac{K_F (K_F -1) (2K_F -1)}{6} + \hbar K_F M
\ee
Clearly the vacua $(K_F , M=K_F )$ and $(K_F +1 , M=0)$ are identical.
Excitations over the Fermi sea come with energies in integer multiples of
$\hbar$ and degeneracies according to the possible fermion arrangements. 

For the bosonized system we choose polar phase space variables,
\be
h_i = \half (p_i^2 + x_i^2 )~,~~~ 
\theta_i = \arctan \frac{x_i}{p_i} ~,~~i=1,2
\ee
in terms of which the single-particle hamiltonian and Poisson structure is
\be
\{ \theta_i , h_j \}\sp = \delta_{ij} ~,~~~ H\sp = h_1 + h_2
\ee
For the droplet description we can take $h_2 = R$ and $\theta_2 = \sigma$
which leaves $( h_1 , \theta_1 ) \sim ( x_1 , p_1 )$
as the residual phase space. The bosonic hamiltonian is
\be
H = \frac{1}{(2\pi\hbar)^2} \int d\sigma dh_1 d\theta_1 (\half R^2 + h_1 R )
\label{Harm}
\ee
The ground state is a configuration with $R + h_1 = E_F$. The nonperturbative
constraints $R>0$, $h_1 >0$ mean that the range of $h_1$ is $0<h_1 <E_F$.

To obtain the matrix representation of $R$ we define oscillator states 
$|a\rangle$, $a=0,1,2,\dots$ in the residual single-particle space $(h_1 , \theta_1 )$
satisfying ${\hat h}_1 |a \rangle = \hbar (a+\half) |a \rangle$. The nonperturbative
constraint for $h_1$ is implemented by restricting to the $K_F$-dimensional
Hilbert space spanned by $a=0,1,\dots K_F$ with $E_F = \hbar K_F -1$.
In the matrix representation $R^{ab}$ becomes a $U(K_F )$ current algebra.
We also Fourier transform in $\sigma$ as in (\ref{Fou}) into discrete modes
$R_n^{ab}$, $n = 0, \pm 1 , \dots$ ($\sigma$ has a period $2\pi$).
The hamiltonian (\ref{Harm})
has no matrix ordering ambiguities (being quadratic in $R$ and $h_1$)
but it needs quantum ordering. Just as in the $1+1$-dimensional case,
we normal order by pulling negative modes $N$ to the left
of positive ones. The result is
\be
\frac{H}{\hbar} = \sum_{n>0} R_{-n}^{ab} R_n^{ba} 
+ \half R_0^{ab} R_0^{ba} + (a+\half) R_0^{aa} 
\label{Hq}
\ee

To analyze the spectrum of (\ref{Hq}) we perform the change of variables
\be
\tR_n^{ab} = R_{n-a+b}^{ab} + (a-K_F +1 ) \delta^{ab} \delta_n
\ee
The new fields $\tR$ satisfy the same Kac-Moody algebra as $R$. The
hamiltonian (\ref{Hq}) becomes
\be
\frac{H}{\hbar} = \sum_{n>0} \tR_{-n}^{ab} \tR_n^{ba} 
+ \half \tR_0^{ab} \tR_0^{ba} + (K_F - \half ) \tR_0^{aa}
+ \frac{K_F (K_F -1)(2K_F -1)}{6} 
\label{Htq}
\ee
The above is the standard quadratic form in $\tR$ plus a constant and
a term proportional to the $U(1)$ charge $\tR_0^{aa} = N - K_F (K_F -1)/2 $.

The ground state consists of the vacuum multiplet $|K_F, M\rangle$,
annihilated by all positive modes $\tR_n$ and transforming in the $M$-fold
fully antisymmetric irrep of $SU(K_F )$ ($0 \le M \le K_F -1$), with
degeneracy equal to the dimension of this representation $K_F ! /M! 
( K_F -M )!$. The $U(1)$ charge of $\tR$ is given by the number of boxes
in the Young tableau of the irreps, so it is $M$. The fermion
number is, then, $N = K_F (K_F -1)/2 +M$. Overall, we have a full
correspondence with the many-body fermion ground states found before;
the state $M = K_F$ is absent, consistently with the fact that the
corresponding many-body state is the state $M=0$ for a shifted $K_F$.

The energy of the above states consists of a constant plus a dynamical
contribution from the zero mode $\tR_0$.
The quadratic part contributes $\half \hbar M$, while the linear part
contributes $\hbar (K_F - \half) M$. Overall, the energy is
$\hbar K_F (K_F -1 ) (2K_F -1)/6 + \hbar K_F M$, also in agreement with
the many-body result.

Excited states are obtained by acting with creation operators $\tR_{-n}$
on the vacuum. These will have integer quanta of energy. Due to the presence
of zero-norm states, the corresponding Fock representation truncates in
just the right way to reproduce the states of second-quantized fermions
with an $SU(K_F )$ internal symmetry and fixed total fermion number.
These particle-hole states are, again, into one-to-one correspondence with
the excitation states of the many-body system, built as towers of 
one-dimensional excited Fermi seas over single-particle states $E_{m,n}$, 
one tower for each value of $n$, with the correct excitation energy.
We have the nonperturbative constraint
$0 \le n < K_F$, as well as constraints related
to the non-depletion of the Fermi sea for each value of $n$, just as
in the one-dimensional case. The number of fermions for each tower can
vary, the off-diagonal operators $\tR_n^{ab}$ creating transitions between
towers, with the total particle number fixed to $N$ by the value of the
$U(1)$ Casimir.

The above will suffice to give a flavor of the \nc bosonization method.
There are clearly many issues that still remain open, not the least of which
is the identification of a fermion creation operator in this framework.
Putting the method to some good use would also be nice.

\section{\foreignlanguage{greek}{T'a p'anta re~i...} (it all keeps flowing...)}

This was a lightning review of the more recent and current aspects of \nc fluids and their uses in many-body
systems. There is a lot more to learn and do. If some of the readers are inspired and motivated into
further study or research in this subject, then this narrative has served its purpose.
We shall stop here.

\end{document}